\documentclass[prd,twocolumn,showpacs,preprintnumbers,amsmath,amssymb,nofootinbib]{revtex4-1}

\usepackage{graphicx}
\usepackage[
colorlinks=true,linkcolor=Blue,citecolor=Blue,urlcolor=BlueViolet,hyperfootnotes=false]{hyperref}
\usepackage{subfigure}
\usepackage{url}
\usepackage{color}
\usepackage{afterpage}

\newcommand{\bbc}{\text{BR}(B_c\to \tau\nu_\tau)}

\newcommand{\dline}[1]{{\parbox{3.5em}{\rule{0cm}{2.3ex}{#1}\strut}}}
\newcommand{\dlinem}[1]{{\parbox{2.5em}{\rule{0cm}{2.3ex}{#1}\strut}}}

\definecolor{BlueViolet}{rgb}{0.2, 0.00, 0.7}
\definecolor{Blue}{rgb}{0.15, 0.00, 0.9}
\definecolor{halayaube}{rgb}{0.4, 0.22, 0.33}
\definecolor{sanddune}{rgb}{0.59, 0.44, 0.09}

\begin{document}

\preprint{PSI--PR--19--09}
\preprint{ZU--TH 26/19}
\preprint{TTP--19--012}
\preprint{P3H--19--011}
\title{\boldmath Addendum to ``Impact of polarization observables and $\boldmath{ B_c\to \tau \nu}$
  on new physics explanations of the $\boldmath {b\to c \tau \nu}$ anomaly''}

\author{Monika Blanke}
\email{monika.blanke@kit.edu}
\affiliation{Institut f\"ur Kernphysik, Karlsruher Institut f\"ur Technologie (KIT), 76344 Eggenstein-Leopoldshafen, Germany\\
 Institut f\"ur Theoretische Teilchenphysik (TTP), Karlsruher Institut f\"ur Technologie (KIT), 76131 Karlsruhe, Germany}

\author{Andreas Crivellin}
\email{andreas.crivellin@cern.ch}
\affiliation{Paul Scherrer Institut, CH--5232 Villigen PSI, Switzerland}
\affiliation{Physik-Institut, Universit\"at Z\"urich, Winterthurerstrasse 190, CH--8057 Z\"urich, Switzerland}

\author{Teppei Kitahara}
\email{teppeik@kmi.nagoya-u.ac.jp}
\affiliation{Institute for Advanced Research, Nagoya University, Furo-cho Chikusa-ku, Nagoya, 464-8602 Japan}
\affiliation{Kobayashi-Maskawa Institute for the Origin of Particles and the Universe, Nagoya University, Nagoya 464-8602, Japan}
\affiliation{Physics Department, Technion--Israel Institute of Technology, Haifa 3200003, Israel}

\author{Marta Moscati}
\email{marta.moscati@kit.edu}
\affiliation{Institut f\"ur Theoretische Teilchenphysik (TTP), Karlsruher Institut f\"ur Technologie (KIT), 76131 Karlsruhe, Germany}

\author{Ulrich Nierste}
\email{ulrich.nierste@kit.edu}
\affiliation{Institut f\"ur Theoretische Teilchenphysik (TTP), Karlsruher Institut f\"ur Technologie (KIT), 76131 Karlsruhe, Germany}
\author{Ivan Ni\v{s}and\v{z}i\'c}
\email{ivan.nisandzic@kit.edu}
\affiliation{Institut f\"ur Theoretische Teilchenphysik (TTP), Karlsruher Institut f\"ur Technologie (KIT), 76131 Karlsruhe, Germany}

\date{May 20, 2019}

\begin{abstract}
In this addendum to Ref.~\cite{Blanke:2018yud}, 
we update our results to include the recent measurement of ${\cal R}(D)$ and ${\cal R}(D^*)$ by the Belle Collaboration~\cite{Abdesselam:2019dgh}: ${\cal R}(D)_{\rm Belle} = 0.307\pm0.037\pm0.016$ and ${\cal R}(D^*)_{\rm Belle} = 0.283\pm0.018\pm0.014$, resulting in the new HFLAV fit result ${\cal R}(D) = {0.340\pm0.027 \pm  0.013}$, ${\cal R}(D^*) = {0.295\pm0.011  \pm 0.008 }$, exhibiting a $3.1\,\sigma$ tension with the Standard Model.
 We present the new fit results and update all figures, including the relevant new collider constraints.
The updated prediction for ${\cal R}(\Lambda_c)$ from our sum rule reads ${\cal R}(\Lambda_c)=
\mathcal{R}_{\rm SM}(\Lambda_c) \left( 1.15 \pm 0.04 \right) =
0.38 \pm 0.01 \pm 0.01$.
We also comment on theoretical predictions for the fragmentation function $f_c$ of $b\to B_c$ and their implication on the constraint from $B_{u/c}\to\tau\nu$ data. 
\end{abstract}
\maketitle

\vspace{-1mm}
\maketitle
\renewcommand{\thefootnote}{\#\arabic{footnote}}
\setcounter{footnote}{0}

In this Addendum, we present an update of our
article~\cite{Blanke:2018yud} in which we studied {the impact} of
polarization observables and the bound on ${\rm BR}(B_c\to \tau\nu)$ on
new physics explanations of the $b\to c\tau\nu$ anomaly.

{Our updated results incorporate} the new experimental results for
$\mathcal{R}(D)$ and $\mathcal{R}(D^\ast)$ measured by the Belle
Collaboration~\cite{Abdesselam:2019dgh}:
\begin{eqnarray}
\begin{aligned}
{\cal R}(D)_{\rm Belle}\,=\,0.307\pm0.037\pm0.016  \,, \label{rdhflav}\\
{\cal R}(D^*)_{\rm Belle}\,=\,0.283\pm0.018\pm0.014\,.
\end{aligned}
\end{eqnarray}
The first quoted error is statistical and the second one is systematic.
The new measurement is consistent with the Standard Model (SM) predictions \cite{Amhis:2016xyh}
\begin{eqnarray}
\begin{aligned}
{\cal R}_{\rm SM}(D)\,=\,0.299\pm0.003 \,, \\
{\cal R}_{\rm SM}(D^*) \,=\,0.258\pm0.005 
\label{eq:SM}
\end{aligned} 
\end{eqnarray}
{at the $0.2\,\sigma$ and $1.1\,\sigma$ level, respectively.}

\begin{figure}[t]
	\begin{center}
		\includegraphics[width=0.48\textwidth,bb= 0 0 511 331]{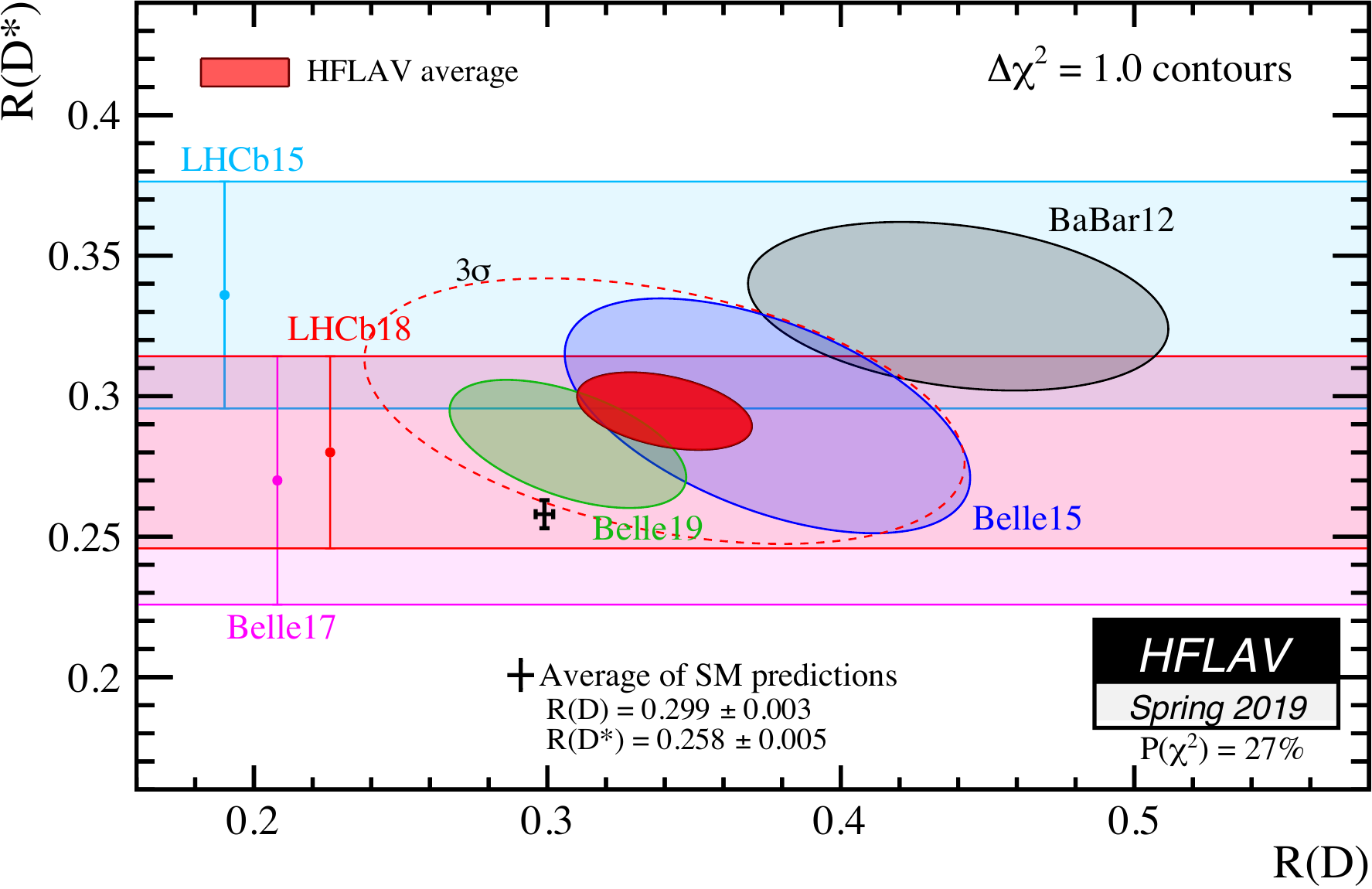}
	\end{center}
	\vspace{-0.5cm}
\caption{The {green}  ellipse shows the result of the new measurement by the Belle Collaboration \cite{Abdesselam:2019dgh}, while the {red} ellipse shows the new world average. The SM predictions are represented by the black bars. {Figure taken from Ref.\ \cite{Amhis:2016xyh}.}}
	\vspace{-0.2cm}
\label{Fig:ellipse}
\end{figure}

Combining this with the previous measurements presented by the BaBar, Belle, and LHCb collaborations in Refs.~\cite{Lees:2012xj,Lees:2013uzd, Huschle:2015rga,Sato:2016svk,Hirose:2016wfn,Hirose:2017dxl,Aaij:2015yra,Aaij:2017uff,Aaij:2017deq},
{the HFLAV Collaboration
\cite{Amhis:2016xyh}
 {has} determined the {averages}}
\begin{eqnarray}
\begin{aligned}
{\cal R}(D)\,=\,{0.340\pm0.027 \pm  0.013}\,, \label{new_average}\\
{\cal R}(D^*)\,=\,{0.295\pm0.011  \pm 0.008 }\,,
\end{aligned}
\end{eqnarray}
with an ${\cal R}(D)$--${\cal R}(D^*)$ correlation of $-0.38$.  {The new
  world averages deviate from the SM at $1.4\,\sigma$
  [${\cal R}(D)$], $2.5\,\sigma$ [${\cal R}(D^*)$], and $3.1\,\sigma$ 
  {[${\cal R}(D)$--${\cal R}(D^*)$ combination]}~\cite{Amhis:2016xyh}}.
{This} situation is shown in Fig.~\ref{Fig:ellipse}.

Including all four observables $\mathcal{R}(D), \mathcal{R}(D^{\ast}), P_\tau(D^\ast)$ and $F_L(D^\ast)$,\footnote{The impact of the $F_L(D^*)$ measurement on new physics in $b\to c\tau\nu$ was previously considered in \cite{Aebischer:2018iyb,Iguro:2018vqb}.} we find the new 
{$p$-value of the two-sided test for the SM 
\begin{align}
p\textrm{-value}_{\rm SM} \sim {0.1}\,\%,
\label{eq:pSM}
\end{align}
which corresponds to a {$3.3\,\sigma$} tension, 
{where we neglect the SM uncertainty.}
Note that our choice of the form factors was explained in
  Ref.~\cite{Blanke:2018yud}, and we obtain the following {central
    values of the SM predictions}:
\begin{eqnarray}
\begin{aligned}
{\cal R}_{\rm SM}(D)\,&=\,0.301\,, ~~~{\cal R}_{\rm SM}(D^*) \,=\,0.254 \,,\\
{P}_{\tau, {\rm SM}}(D)\,&=\,0.32\,, ~~~
{P}_{\tau, {\rm SM}}(D^*) \,=\,-0.49 \,,\\
{F}_{L, {\rm SM}}(D^*) \,&=\,0.46 \,,~~~
{\cal R}_{\rm SM}(\Lambda_c) \,=\,0.33 \,.
\end{aligned}
\end{eqnarray}
{All our fit results are based on these {numbers}.}\footnote{{On the other hand, based on the SM predictions in Eq.~\eqref{eq:SM}, we obtain 
$p$-value$_{\rm SM} \sim 0.2\,\%$ corresponding {to a} $3.1\,\sigma$ tension instead of Eq.~\eqref{eq:pSM}.} }

The authors of Ref.~\cite{Akeroyd:2017mhr} deduced the {stringent}
constraint ${\rm BR}(B_c \to \tau \nu) < 10\%$ from data on a
mixed sample of $B_c^- \to \tau \nu_{\tau}$ and $B^- \to \tau \nu$
candidate events taken at the $Z$ peak in the LEP experiment.  To this
end, the fragmentation function $f_c$ of $b \to B_c^-$ {has been} extracted
from data accumulated at hadron colliders. For asymptotically large
values of the transverse $b$ momentum $p_T$, fragmentation functions are
numbers which are independent of the kinematical variables and the $b$
production mechanism. In Ref.~\cite{Blanke:2018yud}, we pointed out
that hadron collider data {exhibit} a sizable $p_T$ dependence and
{pointed} to production mechanisms beyond fragmentation 
(see also Ref.~\cite{Bardhan:2019ljo}). 
In Fig.~1 of Ref.~\cite{Akeroyd:2017mhr}, $f_c/f_u$ was extracted
from {CMS and LHCb data}. Using the {world average of the} $b\to B^-$
fragmentation function $f_u = 0.404(6)$ \cite{Tanabashi:2018oca}, {we
  find that} the result of Ref.~\cite{Akeroyd:2017mhr} implies
\begin{align}
2.1 \times 10^{-3} \lesssim f_c \lesssim 4.4\times 10^{-3}\,.
\label{fc:LHC}
\end{align}
If one instead uses a calculation of $B_{c}^{-}$ production on the $Z$
peak at $e^{+} e^{-}$ colliders employing nonrelativistic quantum
chromodynamics (NRQCD) at next-to-leading order
\cite{Zheng:2017xgj,Zheng:2019gnb} (see also Ref.~\cite{Kiselev:1994qp}), one
finds
\begin{align}
 f_c \sim 3 \times 10^{-4}\,,
 \end{align}
 with essentially the same estimate for $b \to B_c^{\ast}{}^{-}$
 fragmentation.  If one {further} assumes that $B_c^{\ast}{}^{-}$ decays into
 final states with 
 $B_c^-$ with a branching ratio of $1$,\footnote{{While} $B_c(2S)^{-}$ and
   $B_c^{\ast}(2S)^{-}$ have been observed through a transition of
   $B_c^{(\ast)}(2S)^{-} \to B_c^{(\ast)}{}^{-} \pi^+ \pi^-$
   \cite{Aad:2014laa,Sirunyan:2019osb,Aaij:2019ldo}, no $B_c^{\ast}{}^{-}$ has
   been detected yet.} then $f_c$ effectively changes to
 \begin{align}
 f_c \sim 6 \times 10^{-4}\,.
 \label{fc:Z}
\end{align}
Therefore by  comparing Eqs.~\eqref{fc:LHC} and \eqref{fc:Z}, {we 
conclude} that the constraint on ${\rm BR}(B_c\to \tau \nu)$ derived 
in Ref.~\cite{Akeroyd:2017mhr} is too stringent by a factor of 3 to 4.
Taking into account the intrinsic uncertainties of the NRQCD
calculation, the $Z$ peak data cannot rule out our most conservative 
scenario which permits ${\rm BR}(B_c\to \tau \nu)$ to be as large as 60\%.

\begin{table*}[th]
	\begin{tabular}{|c||c|c|c|c|c||c|c|c|c|c|c|c|}\hline
1D hyp.   & best-fit & $1\,\sigma$ range & $2\,\sigma$ range & $p$-value (\%)
& pull$_{\rm SM}$  & ${\cal R}(D)$ & ${\cal R}(D^*)$& $F_L({D^{*}})$ & $P_\tau (D^*)$ & $P_\tau (D)$ & ${\cal R}(\Lambda_c) $\\
\hline\hline    		
$C_V^L$    &  0.07 &   [0.05, 0.09]  & [0.04, 0.11] &44&  4.0   &
\dline{0.347  $+0.2\,\sigma$} &  \dline{0.292  $-0.2\,\sigma$}    &  \dline{0.46  $-1.6\,\sigma$}   & \dline{$-0.49$  $-0.2\,\sigma$}& \dline{0.32 \\}&\dline{0.38 \\}\\
\hline 		
$C_S^R$    & 0.09  &   [0.06, 0.11]   &  [0.03, 0.14]  & 2.7 &  3.1 & \dline{0.380 $+1.4\,\sigma$} & \dline{0.260 $-2.6\,\sigma$}& \dline{0.47 $-1.5\,\sigma$}& \dline{$-0.46$  $-0.1\,\sigma$}& \dline{0.46 \\}&\dline{0.36 \\}\\
\hline  
$C_S^L$    &  0.07 &  [0.04, 0.10]    &  $[-0.00, 0.13]$  &0.26 &  2.1   &  \dline{0.364 $+0.8\,\sigma$}& \dline{0.250 $-3.3\,\sigma$}&   \dline{0.45 $-1.7\,\sigma$}  &\dline{$-0.51$ $-0.2\,\sigma$}& \dline{0.44 \\ }&\dline{0.35 \\ }\\
\hline
\scalebox{.9}{$C_S^L=4C_T$} & $-0.03$  &   [$-0.07$, $0.01$]   &  [$-0.11$, 0.04]  & 0.04& 0.7    & 
\dline{0.278 $-2.1\,\sigma$}& \dline{0.263 $-2.3\,\sigma$}  & \dline{0.46 $-1.6\,\sigma$}     &\dline{$-0.47$ $-0.2\,\sigma$} & \dline{0.27 \\}&\dline{0.33 \\}\\
\hline
\end{tabular}
\caption{Updated fit results for the 1D hypotheses (hyp.)\ of Ref.~\cite{Blanke:2018yud}, with the Wilson coefficients defined at the scale $\mu=1\,$TeV. 
}
	\label{tab:results1D}
\end{table*}

\begin{table*}[th]
\begin{tabular}{|c||c|c|c||c|c|c|c|c|c|c|c|c|}\hline
  	2D hyp.  & best-fit& $p$-value {(\%)} & pull$_{\rm SM}$  & ${\cal R}(D)$ & ${\cal R}(D^*)$ & $F_L({D^{*}})$ &  $P_\tau (D^*)$ & $P_\tau (D)$ & ${\cal R}(\Lambda_c) $\\\hline\hline
  	$(C_V^L,\,C_S^L=-4C_T)$&($0.10,-0.04)$&29.8&3.6&
  	\dline{0.333 $-0.2\,\sigma$}& \dline{0.297 $+0.2\,\sigma$}&\dline{0.47 $-1.5\,\sigma$} &\dline{$-0.48$ $-0.2\,\sigma$}& \dlinem{0.25 \\}& \dlinem{0.38 \\}\\ \hline
  	$\left(C_S^R,\,C_S^L)\right|_{60 \%}$& \parbox{7em}{\rule{0pt}{2.3ex}($0.29,-0.25)$ $(-0.16,-0.69)$\strut}&75.7&3.9&
  	\dline{0.338 $~~~0.1\,\sigma$} &\dline{0.297 $+0.1\,\sigma$}& \dline{0.54 $-0.7\,\sigma$}&\dline{$-0.27$ $+0.2\,\sigma$}& \dlinem{0.39 \\}&\dlinem{0.38 \\}\\ \hline
  	$\left(C_S^R,\,C_S^L)\right|_{30 \%}$&\parbox{7em}{\rule{0pt}{2.3ex}($0.21,-0.15)$ $(-0.26,-0.61)$\strut}&30.9&3.6&
  	\dline{0.353 $+0.4\,\sigma$}&\dline{0.280 $-1.1\,\sigma$}&\dline{0.51  $-1.0 \,\sigma$}& \dline{$-0.35$ $~~~0.0\,\sigma$}&\dlinem{0.42 \\}&\dlinem{0.37\\}\\ \hline
  	$\left(C_S^R,\,C_S^L)\right|_{10 \%}$
  	&\parbox{7em}{\rule{0pt}{2.3ex}($0.11,-0.04)$ $(-0.37,-0.51)$\strut}&2.6&2.9&\dline{0.366 $+0.9\,\sigma$} &\dline{0.263 $-2.3\,\sigma$}&\dline{0.48 $-1.4\,\sigma$} &\dline{$-0.44$ $-0.1\,\sigma$} &\dlinem{0.44 \\}&\dlinem{0.36\\}\\ \hline 
  	$(C_V^L,\,C_S^R)$&($0.08,-0.01)$&26.6&3.6&
  	\dline{0.343 $+0.1\,\sigma$}&\dline{0.294 $-0.1\,\sigma$}&\dline{0.46 $-1.6\,\sigma$} &\dline{$-0.49$ $-0.2\,\sigma$}&\dlinem{0.31 \\}&\dlinem{0.38 \\}\\ \hline 
  	\scalebox{.86}{$\left.({\rm Re}[C_S^L=4C_T],\,{\rm Im}[C_S^L=4C_T])\right|_{60,30\%}$}&($-0.06,\pm 0.31)$&25.0&3.6&
  	\dline{0.339 $0.0\,\sigma$}&\dline{0.295 0.0 $\,\sigma$}
  	&\dline{0.45 $-1.7\,\sigma$} 
  	&\dline{$-0.41$ $-0.1\,\sigma$}&\dlinem{0.41 \\}&\dlinem{0.38 \\}
  	\\ \hline
  	\scalebox{.87}{$\left. ({\rm Re}[C_S^L=4C_T],\,{\rm Im}[C_S^L=4C_T])\right|_{10\%}$} &($-0.03,\pm 0.24)$&5.9&3.2&
  	\dline{0.330 $-0.3\,\sigma$}& \dline{0.275 $-1.4\,\sigma$}&\dline{0.46  $-1.6\,\sigma$}& \dline{$-0.45$ $-0.1\,\sigma$} &\dlinem{0.38 \\}&\dlinem{0.36 \\}\\ \hline
\end{tabular}
 \caption{Updated fit results for the 2D hypotheses (hyp.)\ of Ref.~\cite{Blanke:2018yud}, with the Wilson coefficients defined at the scale $\mu=1\,$TeV. }	
  	\label{tab:results2D}
  \end{table*}
  
\begin{figure*}[t]
	\begin{center}
		\includegraphics[width=0.455\textwidth, bb= 0 0 260 143]{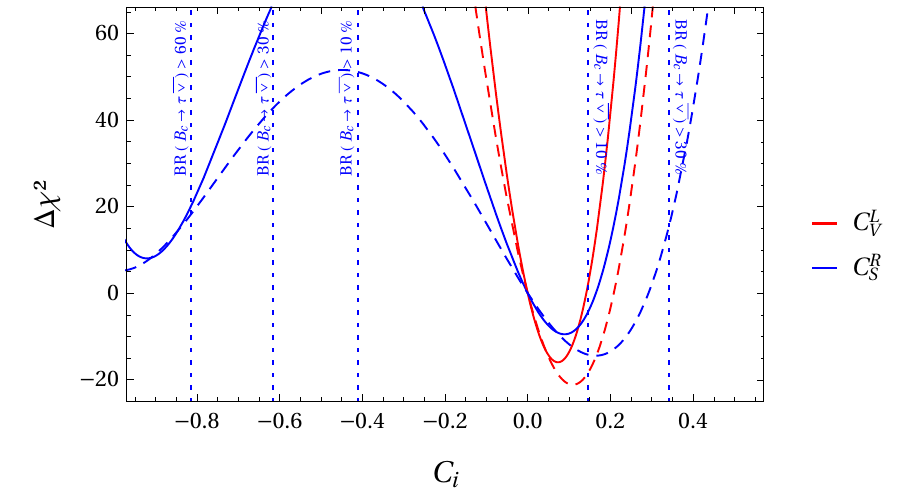}
				\includegraphics[width=0.505\textwidth, bb= 0 0 260 130]{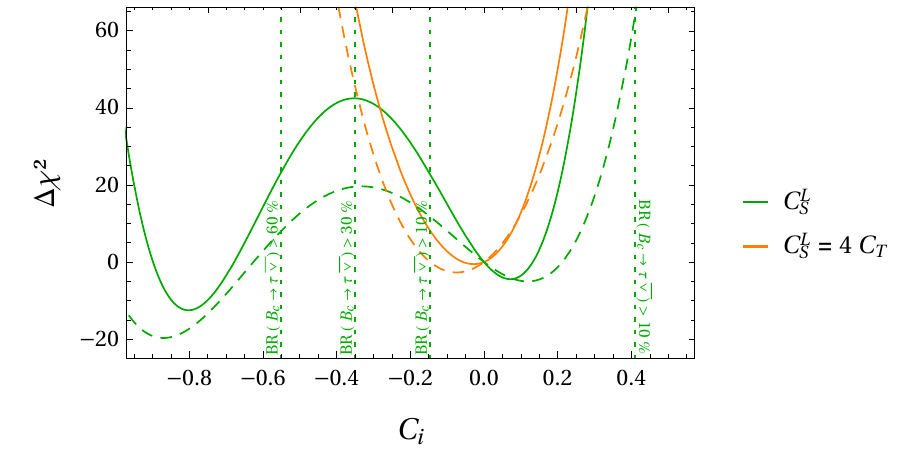}
	\end{center}
	\vspace{-0.5cm}
\caption{$\Delta \chi^2$ {of $\mathcal{R}(D), \mathcal{R}(D^{\ast}), P_\tau(D^\ast)$ and $F_L(D^\ast)$} for the four one-dimensional (1D) scenarios where $\mu = 1\,$TeV. 
{The dashed lines do not include the latest Belle results \cite{Abdesselam:2019dgh}, while the solid lines include {all} data.}
The dotted vertical lines correspond to the limit on $C_S^{L,R}$ from ${\rm BR}(B_c\to \tau \nu)$ assuming a maximal value of $10\%$, $30\%$ or $60\%$.
Best-fit points are not constrained from the $10\%$ limit.
}
	\label{WCsingle}
\end{figure*}

\begin{figure*}[th]
\begin{center}
{
\includegraphics[width=0.4\textwidth, bb= 0 0 260 248]{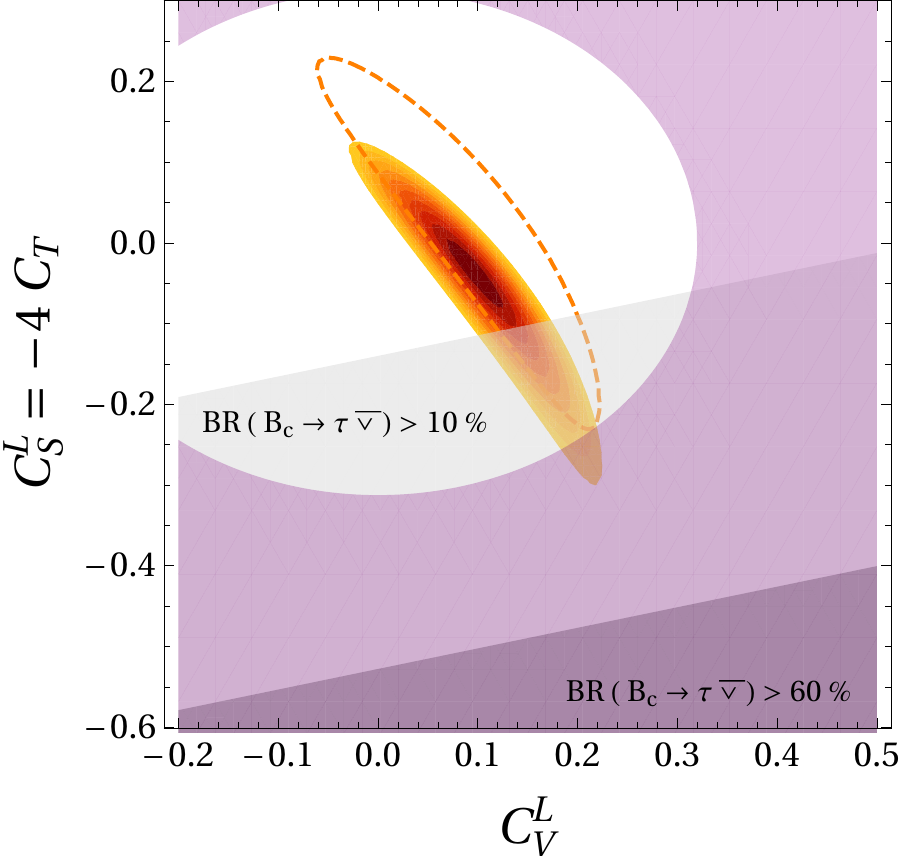}
}
{
\includegraphics[width=0.47\textwidth, bb=0 0 304 253]{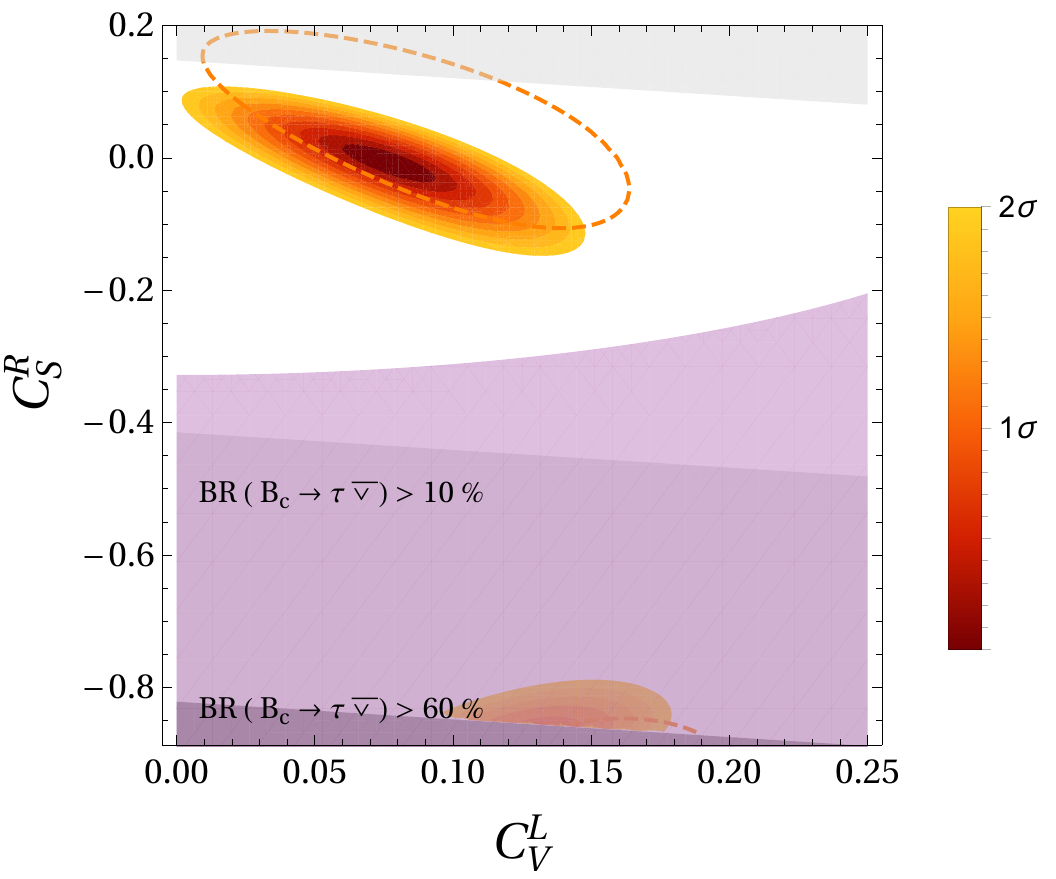}
}
\\
\vspace{0.4cm}
\hspace{0.5cm}
{
\includegraphics[width=0.4\textwidth, bb= 0 0 260 249]{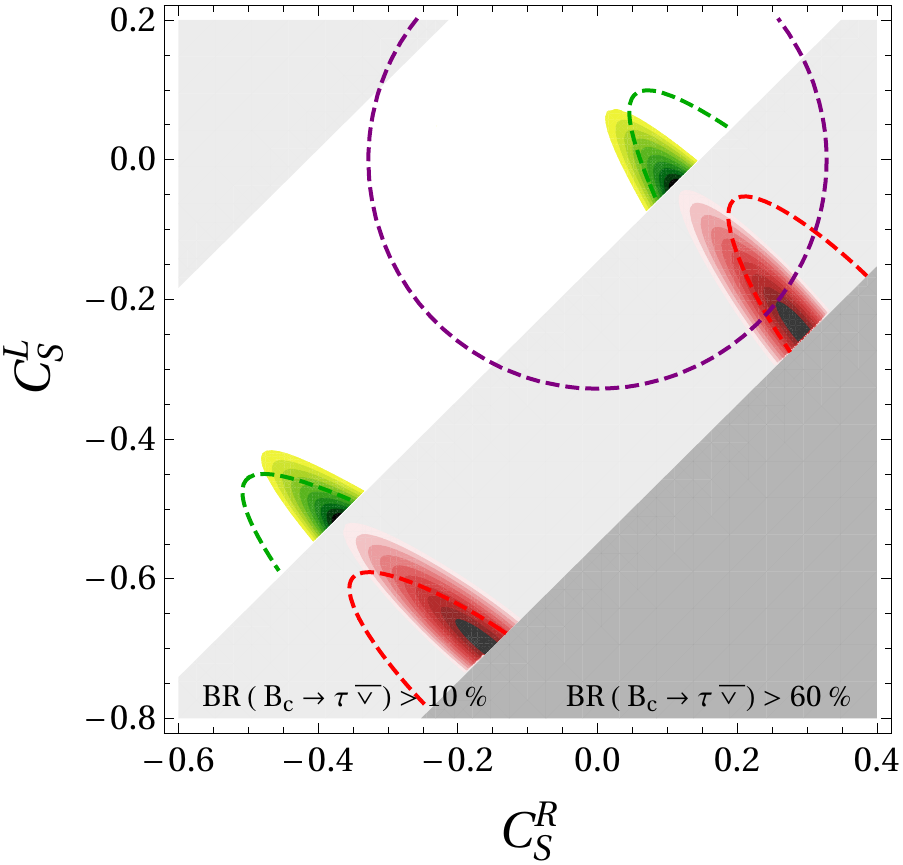} 
}
{
\includegraphics[width=0.51\textwidth, bb=0 0 336 251]{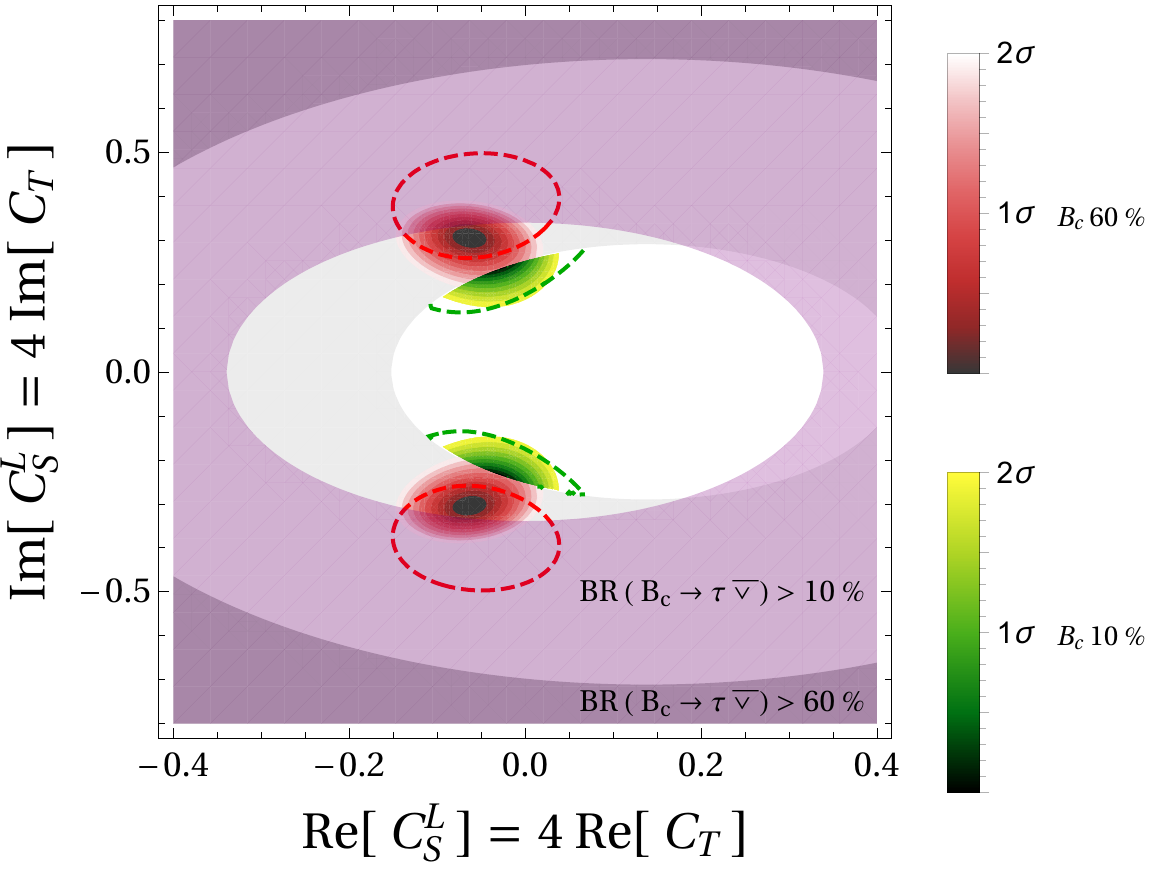}
}
\caption{Updated results of the fits for the $2\,\sigma$ regions in the four 2D scenarios of Ref.~\cite{Blanke:2018yud}, 
with Wilson coefficients given at the matching scale of $1\,$TeV.
{The dashed contours do not include the latest Belle results \cite{Abdesselam:2019dgh}, while the shaded ellipses include {all} data.}
The current collider bounds in Eq.~\eqref{eq:collider} exclude the purple shaded regions at the $2\,\sigma$ level.
The dashed purple circle in the lower left plot indicates the collider constraint on the charged Higgs scenario (see text). }
\label{WCdouble}
\end{center}
\end{figure*}

Tables \ref{tab:results1D} and \ref{tab:results2D} update the respective tables in  Ref.~\cite{Blanke:2018yud}, showing the numerical results of the fit in the various one- (1D) and two-dimensional (2D) scenarios for the Wilson coefficients. The corresponding plots are shown in Figs.~\ref{WCsingle} and \ref{WCdouble}. In all cases, the best-fit points moved closer to the SM, with the biggest change being in the one-dimensional scalar scenarios. In the $C_S^R$ scenario, the best-fit point is hence no longer in tension with the aggressive $\text{BR}(B_c\to\tau\nu)<10\%$ bound.

The most general and powerful collider constraint on the $b\to c\tau\nu$ operators comes from 
 high-$p_T$ tails in mono-$\tau$ searches.
Reference~\cite{Greljo:2018tzh} investigated the constraints on the effective field theory (EFT) operators mediating $b\to c\tau\nu$.
This EFT
  analysis is valid for certain leptoquark models if the leptoquarks are
  sufficiently heavy.\footnote{Direct searches for leptoquarks coupled to third-generation quarks constrain their masses to roughly $m_\text{LQ} > 1\,\text{TeV}$ \cite{Angelescu:2018tyl,Sirunyan:2018ruf}. 
  These direct collider bounds significantly depend on the branching fractions of the leptoquarks.}  
The {resulting} $2\,\sigma$ upper bounds from the current collider data are~\cite{Greljo:2018tzh} 
\begin{align}
\left| C_V^L \right|< 0.32\,, ~\left|C_S^{L(R)}\right|<0.57\,,~\left|C_T\right| < 0.16\,,
\label{eq:collider}
\end{align}
at the scale $\mu = m_b$.
 In Fig.~\ref{WCdouble}, we apply these collider bounds to the four two-dimensional scenarios, where we assume that
interference between two different operators is suppressed. Note that in contrast to our findings in Ref.~\cite{Blanke:2018yud}, the best-fit points in the complex $C_S^L = 4 C_T$ scenario are no longer in tension with the collider constraints.
{Scenarios with color-singlet $s$-channel mediators, like a charged scalar,
  require} model-dependent studies beyond the EFT framework, see e.\,g.\  Ref.\ \cite{Faroughy:2016osc,Iguro:2018fni}.
Hence, for the $(C_S^R,\,C_S^L)$ scenario originating from the exchange of a charged Higgs boson,  the collider bound is valid only in the heavy-mass limit, and we therefore indicate it by a dashed line.

Figure \ref{Fig:Correl-First} shows the prediction for ${\cal R}(\Lambda_c)$ in the four two-dimensional scenarios, as functions of ${\cal R}(D)$ and ${\cal R}(D^*)$, respectively. 
{In Ref.~\cite{Blanke:2018yud}, we obtained a sum rule}
\begin{eqnarray}
\frac{\mathcal{R}(\Lambda_c)}{\mathcal{R}_{\rm SM}(\Lambda_c)}
  \,\simeq\, 0.262 \frac{\mathcal{R}(D)}{\mathcal{R}_{\rm SM}(D)} + 0.738 \frac{\mathcal{R}(D^*)}{\mathcal{R}_{\rm
      SM}(D^*)} \,. \label{eq:sumrule}
\end{eqnarray}
The decrease in ${\cal R}(D^{(*)})$ implied by the new Belle measurement leads to a decreased prediction for $ \mathcal{R}(\Lambda_c)$ through our sum rule \cite{Blanke:2018yud} 
\begin{eqnarray}
\begin{aligned}
 \mathcal{R}(\Lambda_c) \,=\,& \mathcal{R}_{\rm SM}(\Lambda_c) \left( 1.15 \pm 0.04 \right) 
  \label{eq:predlc1} \\
  \,=\,&  0.38 \pm 0.01 \pm 0.01      \label{eq:predlc2}\,,
\end{aligned} 
\end{eqnarray}
where the first error arises from the experimental uncertainty of $\mathcal{R}(D^{(*)})$, 
while the second error comes from the form factors.
{This model-independent relation between $\mathcal{R}(D)$, $\mathcal{R}(D^*)$, and $\mathcal{R}(\Lambda_c)$ originates from heavy-quark symmetry:~in the heavy-quark limit the inclusive $b\to c\tau\nu$ rate is saturated by the sum of $B\to D\tau\nu$ and $B\to D^*\tau\nu$ in the mesonic case, and by $\Lambda_b\to\Lambda_c\tau\nu$ in the baryonic case \cite{Mannel_private}.} {
We have checked that the sum rule in Eq.~\eqref{eq:sumrule} also holds for new physics scenarios with right-handed neutrinos, although they are not considered in our analysis.}

\begin{figure*}[tp]
	\subfigure{
		\includegraphics[width=0.39\textwidth, bb= 0 0 350 294]{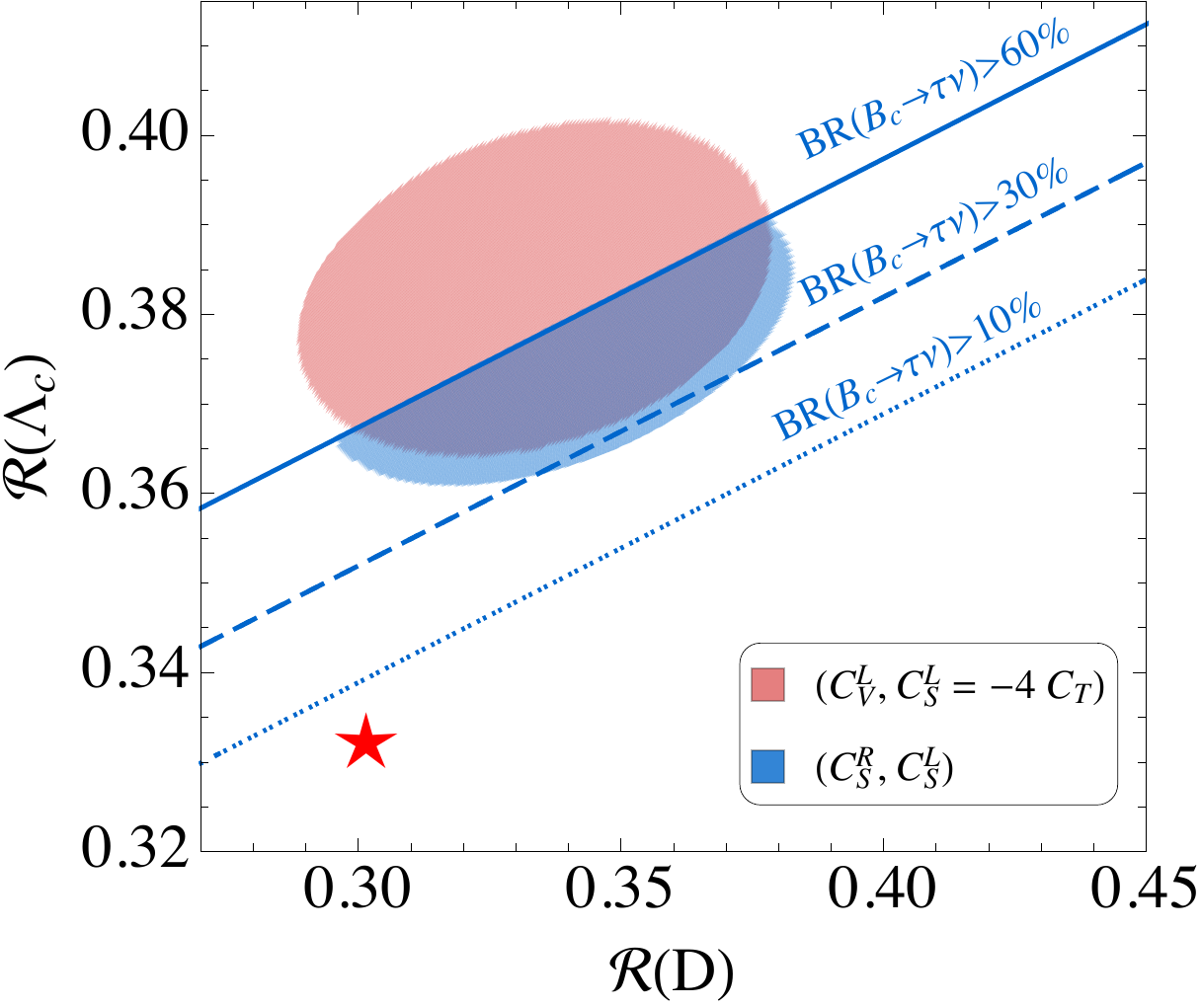}}\quad
	\subfigure{	
		\includegraphics[width=0.405\textwidth, bb= 0 0 350 294]{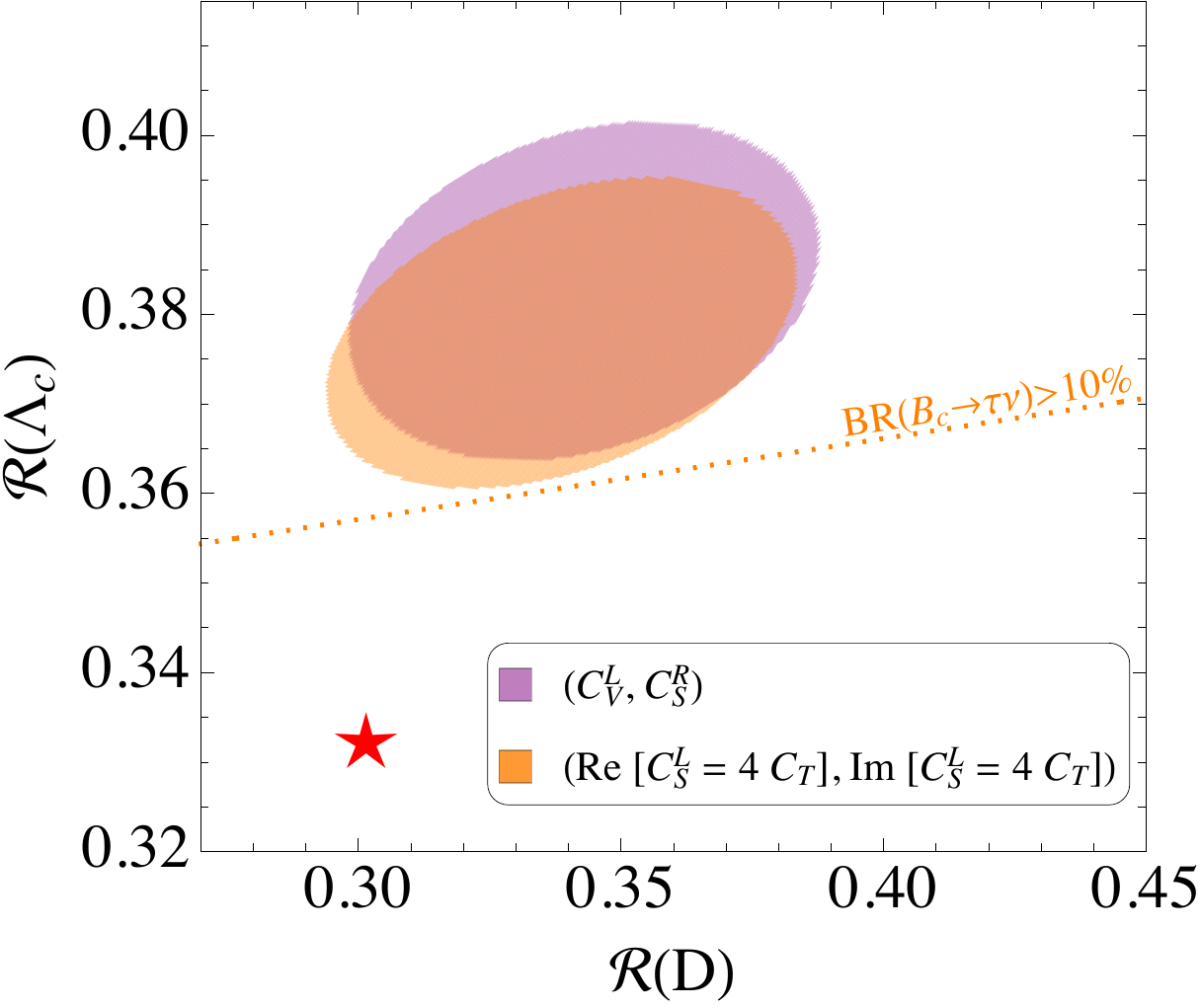}          }\\
	\subfigure{		\includegraphics[width=0.40\textwidth, bb= 0 0 350 304]{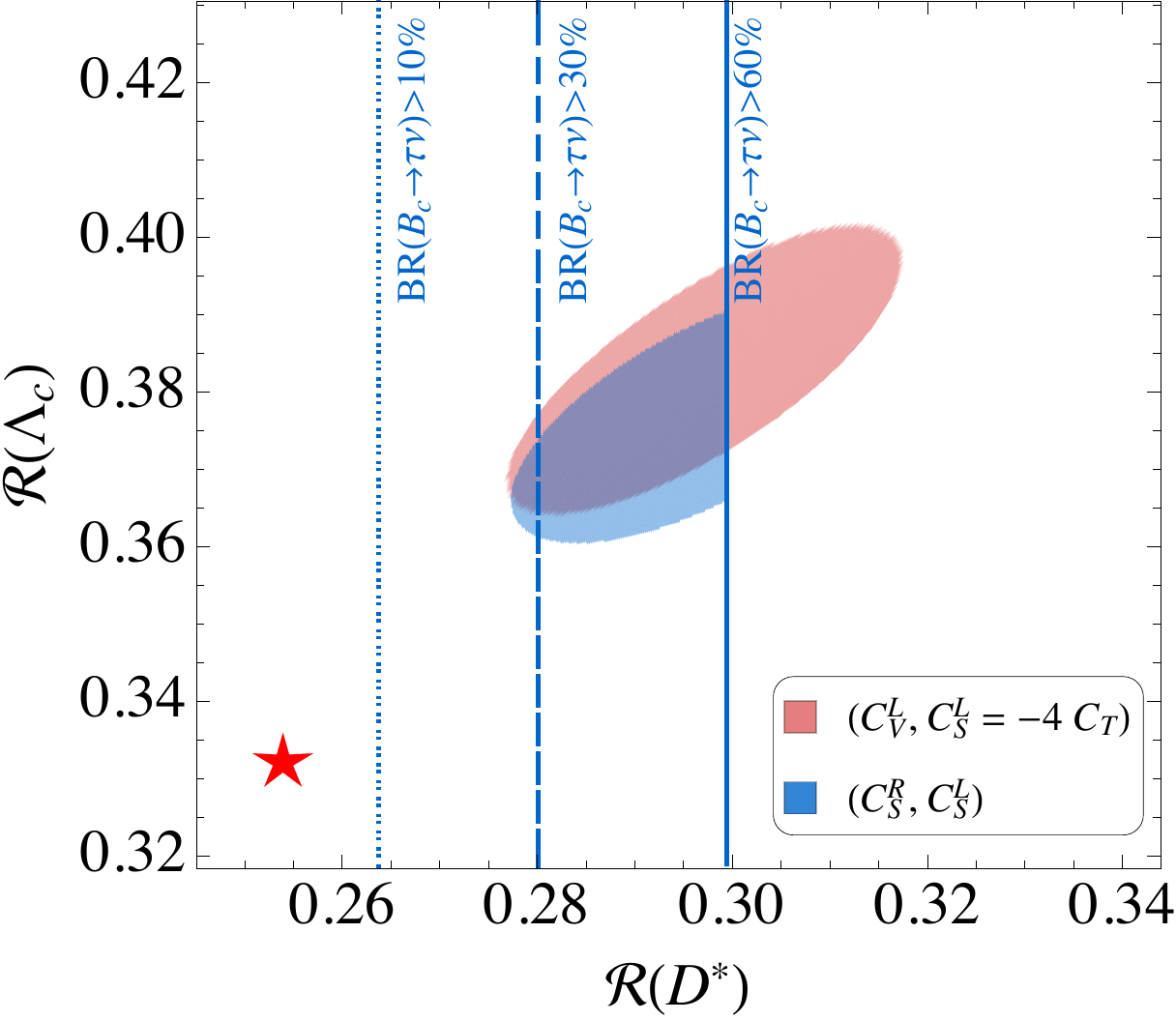}			}\quad
	\subfigure{	
		\includegraphics[width=0.38\textwidth, bb= 0 0 350 304]{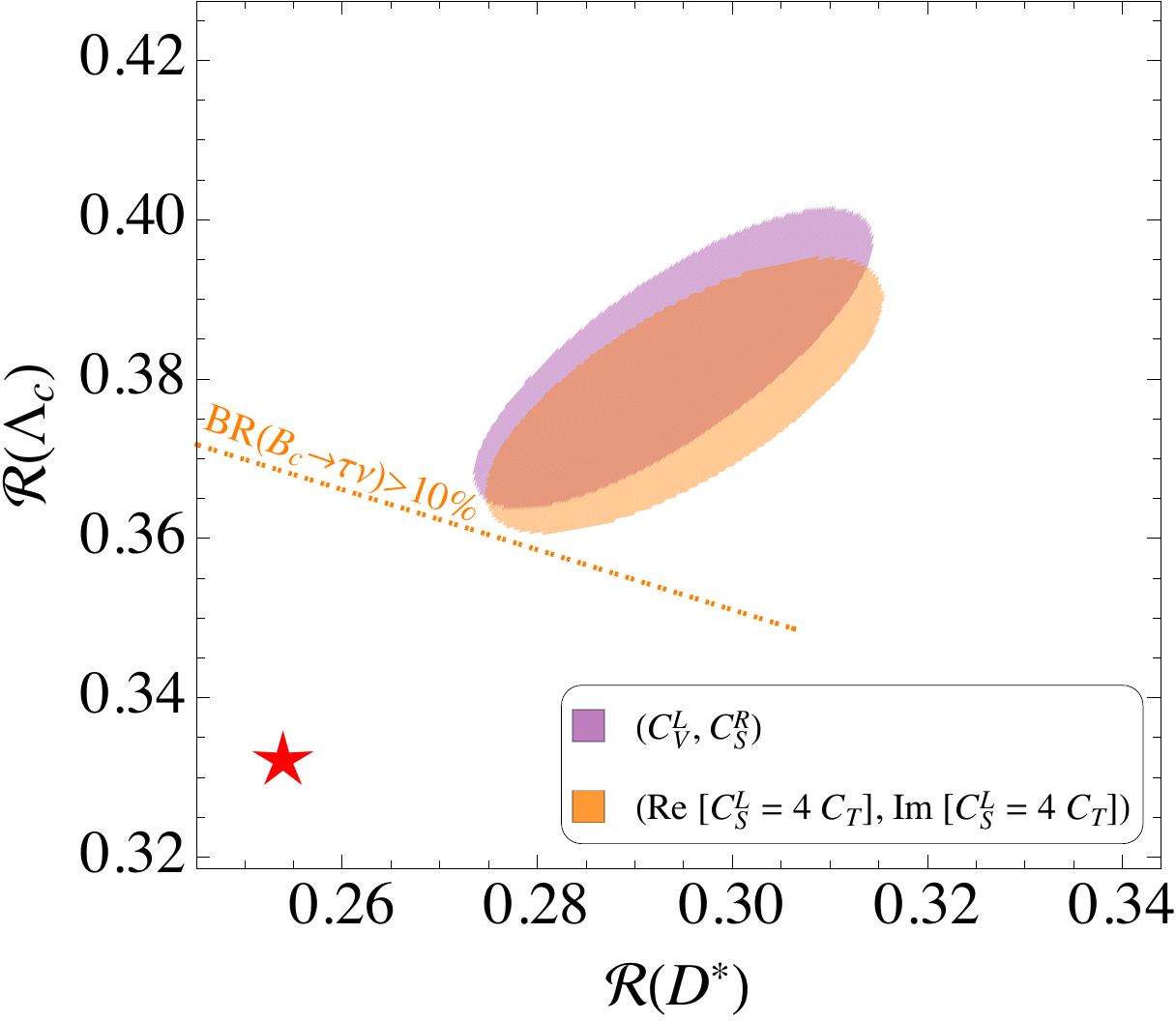}          }\\
\caption{Preferred $1\,\sigma$ regions in the four two-dimensional scenarios in the $\mathcal{R}(D^{(\ast)})$--$\mathcal{R}(\Lambda_c)$ plane for $\bbc<60\%$, updating Fig.~3 of Ref.~\cite{Blanke:2018yud}.}
\label{Fig:Correl-First}
\end{figure*}

As shown in Fig.~\ref{Fig:Correl-Second}, the pairwise correlations between the polarization observables $P_\tau(D)$, $P_\tau(D^*)$, and $F_L(D^*)$ are still distinct for the various two-dimensional scenarios. In order to fully exploit their potential, besides better measurements more precise theoretical predictions for the $B\to D$ and $B\to D^*$ form factors are also necessary.

\begin{figure*}[tp]
	\subfigure{
		\includegraphics[width=0.40\textwidth, bb= 0 0 350 302]{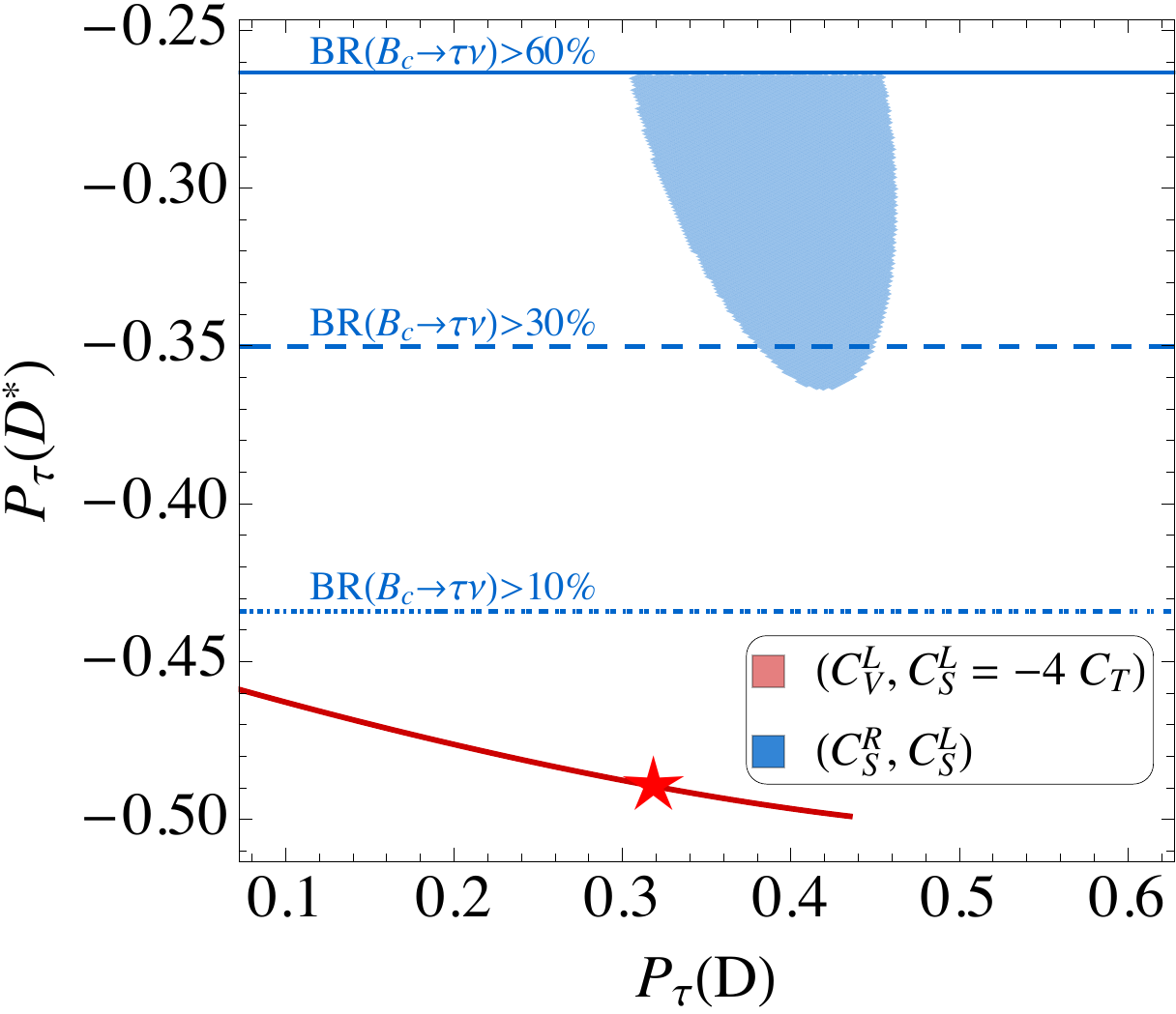}			}\quad
	\subfigure{	
		\includegraphics[width=0.40\textwidth, bb= 0 0 350 297]{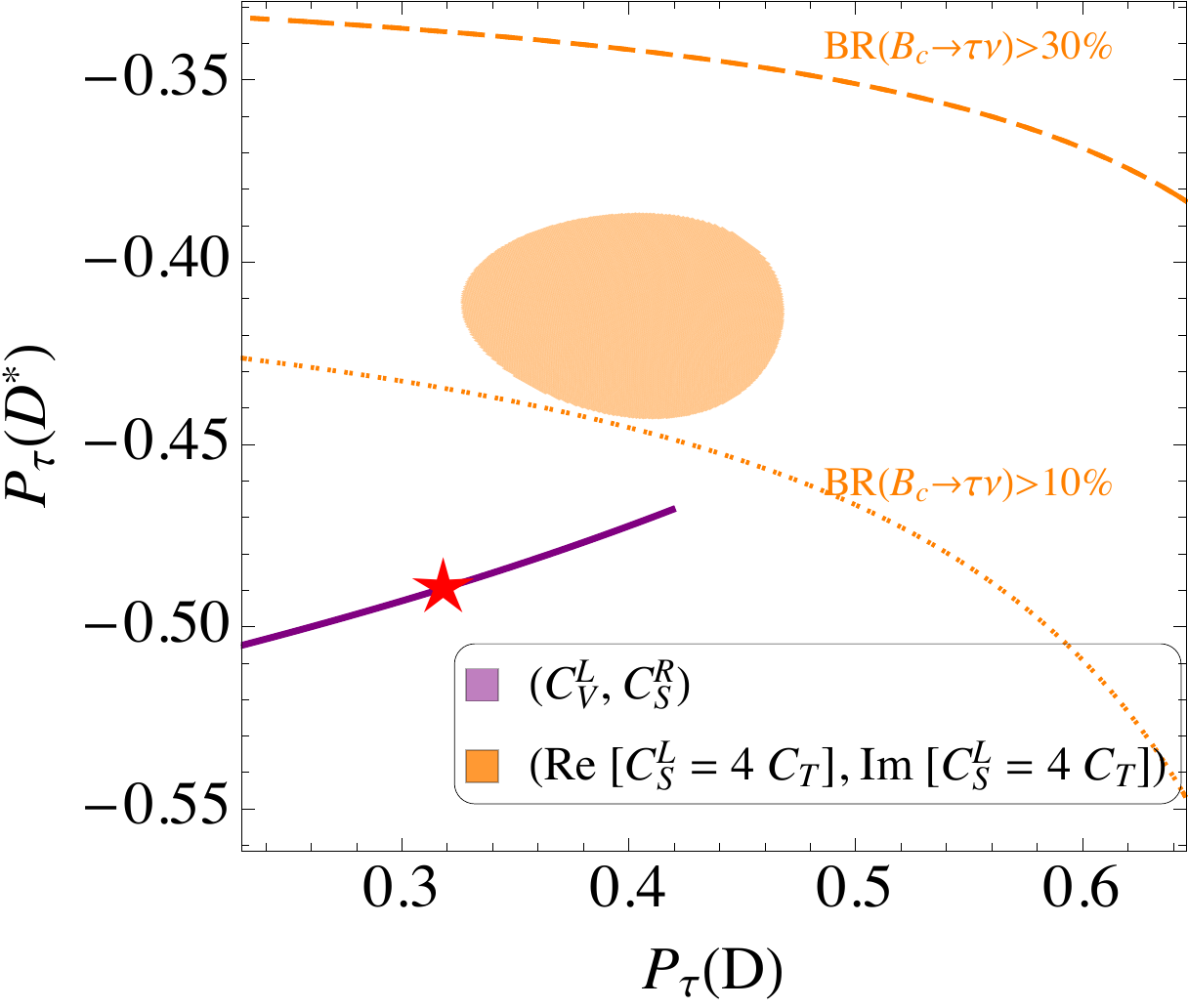}	        }\\
	\subfigure{
		\includegraphics[width=0.41\textwidth, bb=0 0  350 301]{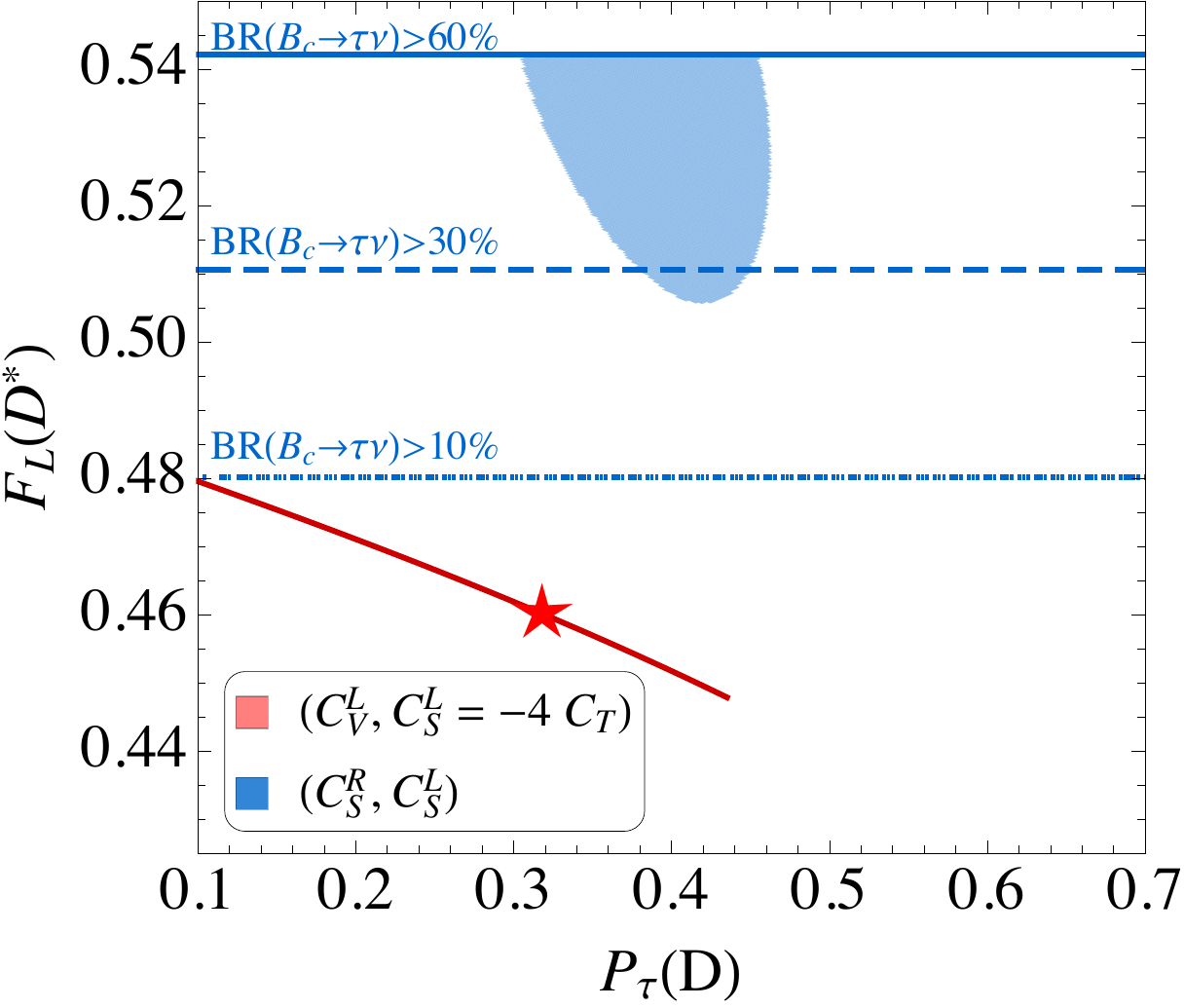}			}\quad
	\subfigure{	
		\includegraphics[width=0.39\textwidth, bb= 0 0 350 317]{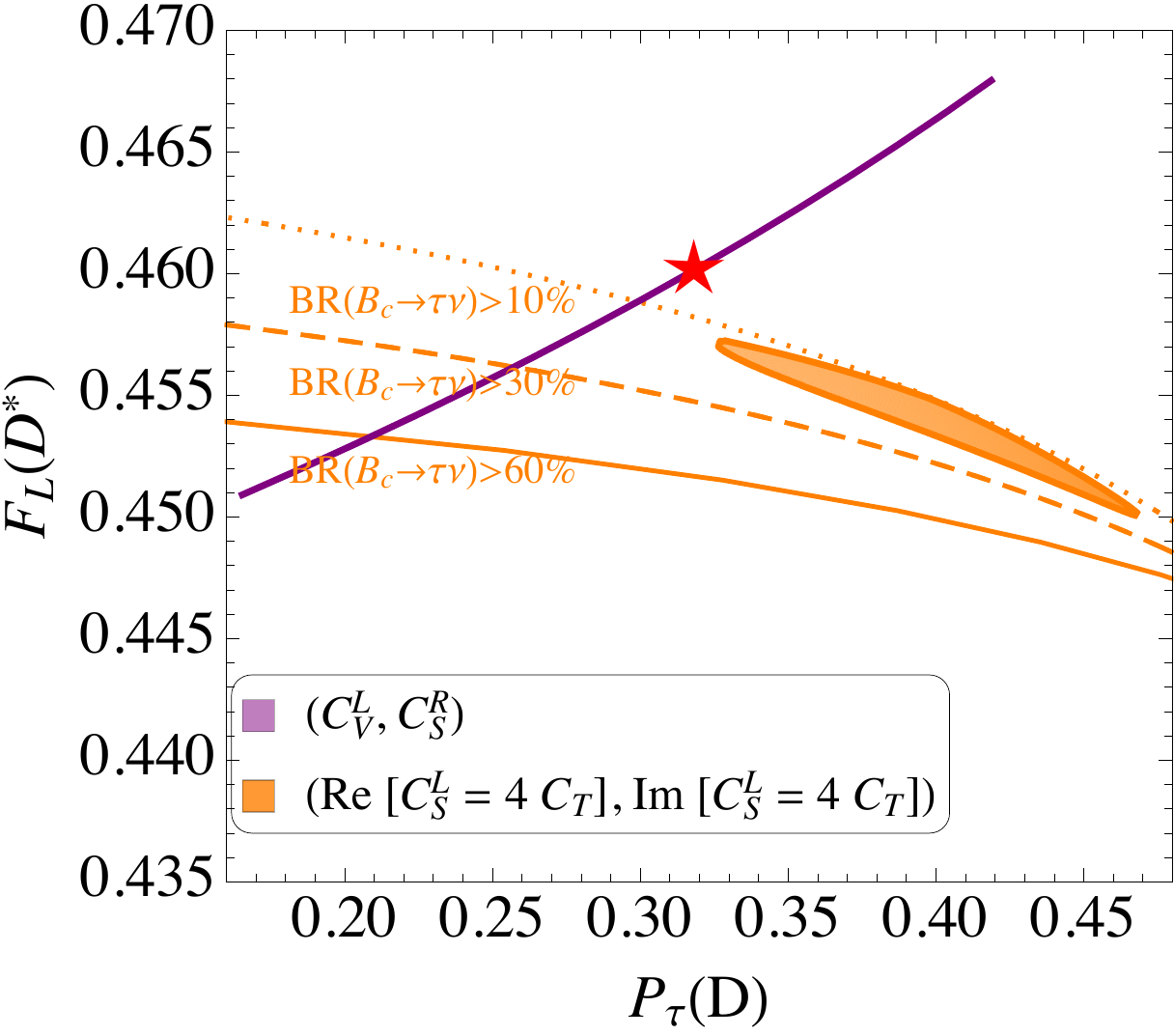}        }\\
	\subfigure{
		\includegraphics[width=0.41\textwidth, bb = 0 0 350 298]{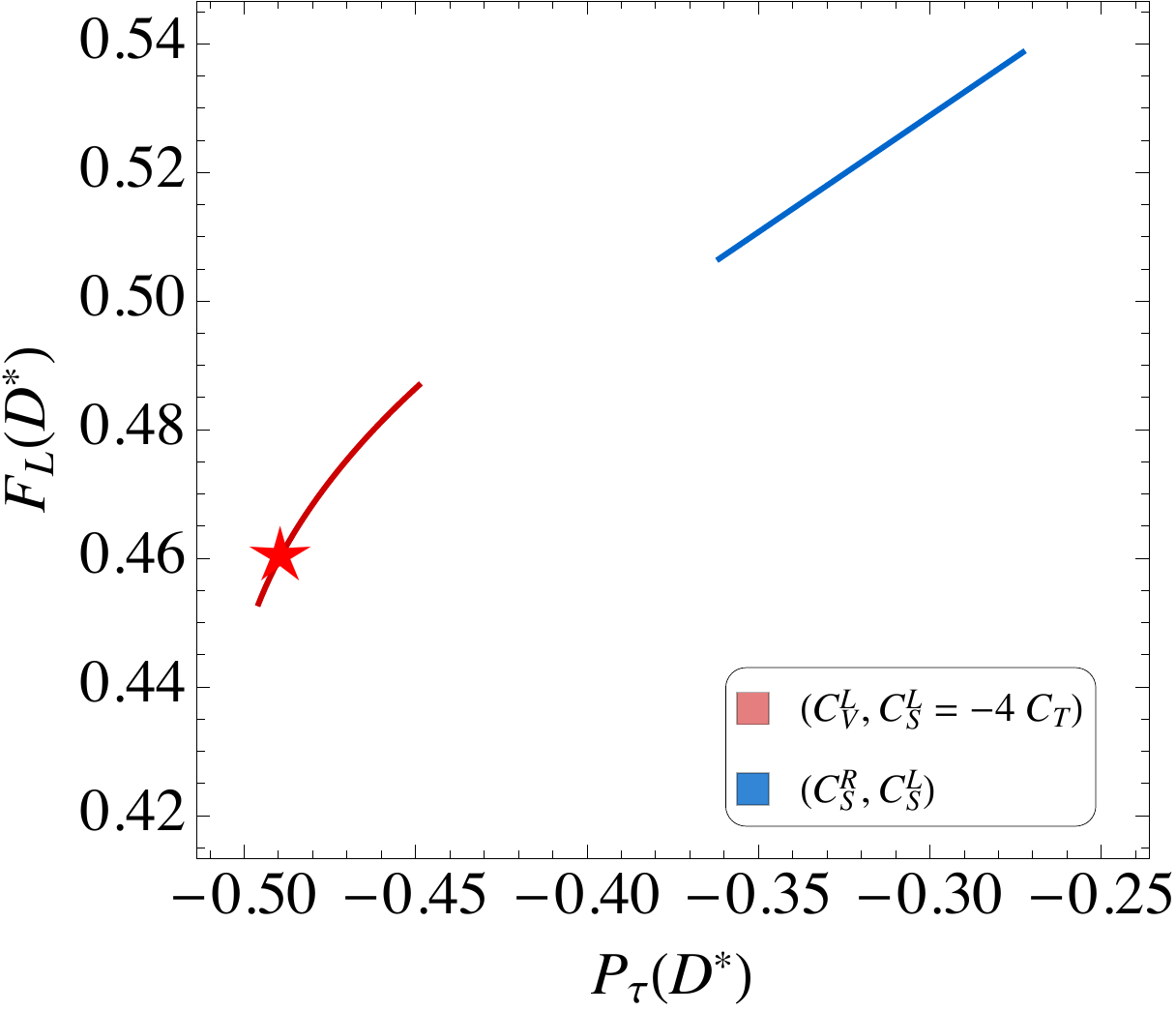}			}\quad
			\subfigure{	
		\includegraphics[width=0.40\textwidth, bb= 0 0 350 309]{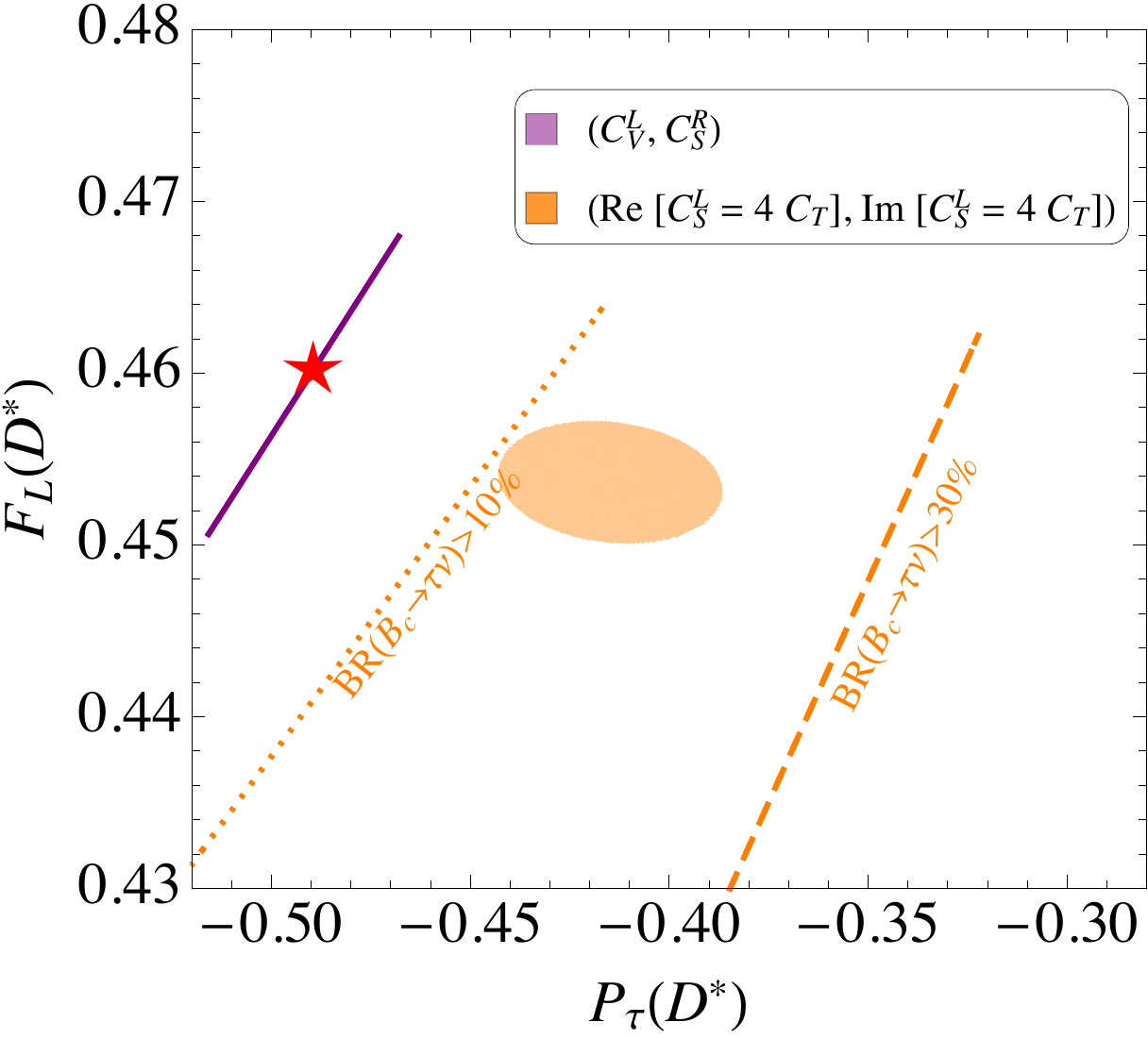}        }\\
\caption{Pairwise correlations between the observables
$P_\tau (D), P_\tau (D^\ast)$ and $F_L(D^\ast)$, updating Fig.~4 of Ref.~\cite{Blanke:2018yud}.}
\label{Fig:Correl-Second}
\end{figure*}

Figures~\ref{RDRDstar1} and \ref{RDRDstar2} show the contour lines of the polarization observables $P_\tau(D)$, $P_\tau(D^*)$, and $F_L(D^*)$ and the ratio ${\cal R}(\Lambda_c)$ in the  ${\cal R}(D)$--${\cal R}(D^*)$ plane. In these plots only the position of the experimentally preferred region for  ${\cal R}(D)$ and ${\cal R}(D^*)$ has been changed with respect to the version shown in Figs.~5 and 6 of Ref.~\cite{Blanke:2018yud}.

\begin{figure*}[tp]
\begin{center}
	\subfigure{
		\includegraphics[width=0.45\textwidth, bb=0 0 360 360]{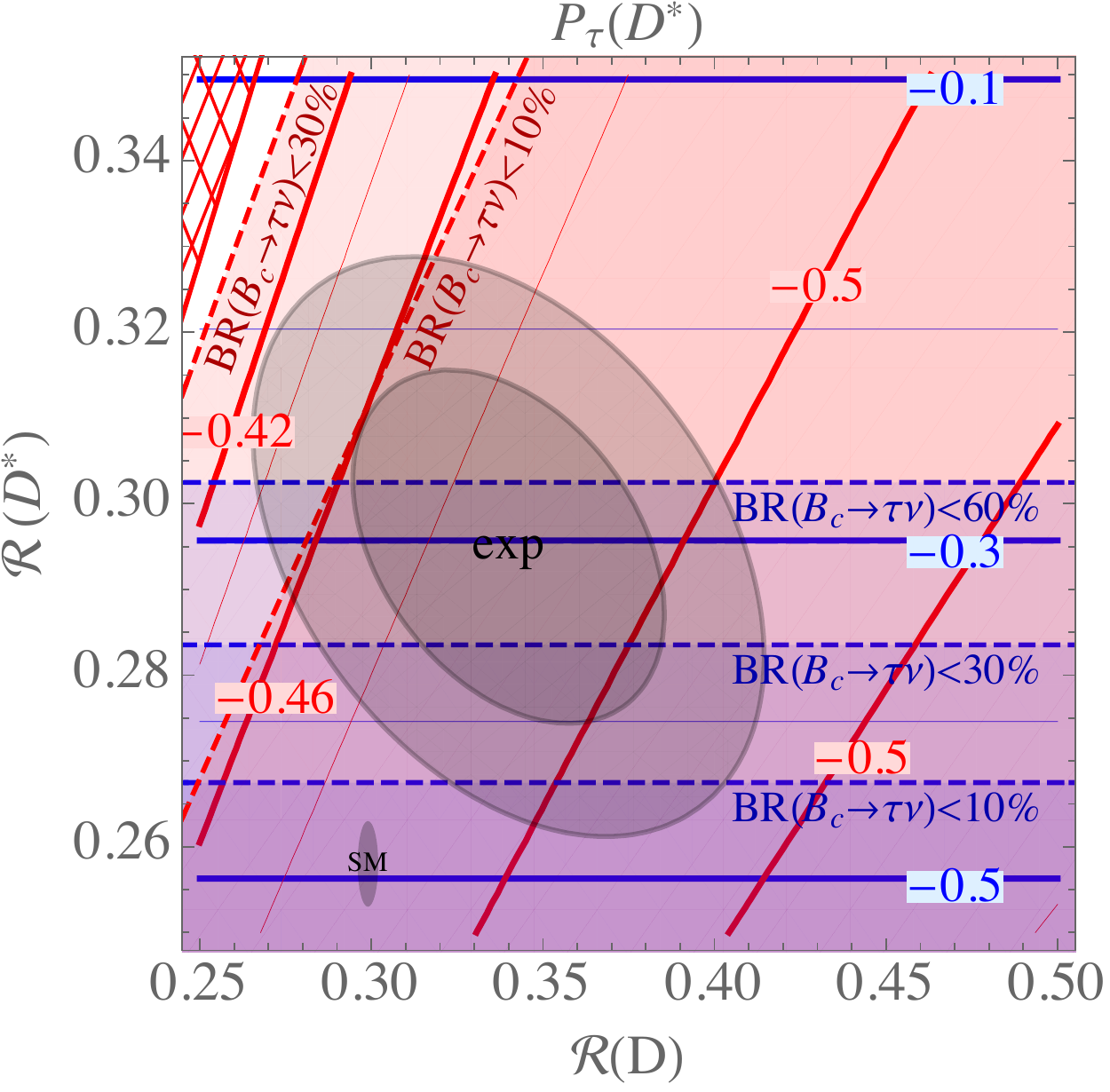}			}\qquad
	\subfigure{		
		\includegraphics[width=0.45\textwidth, bb =0 0  360 360]{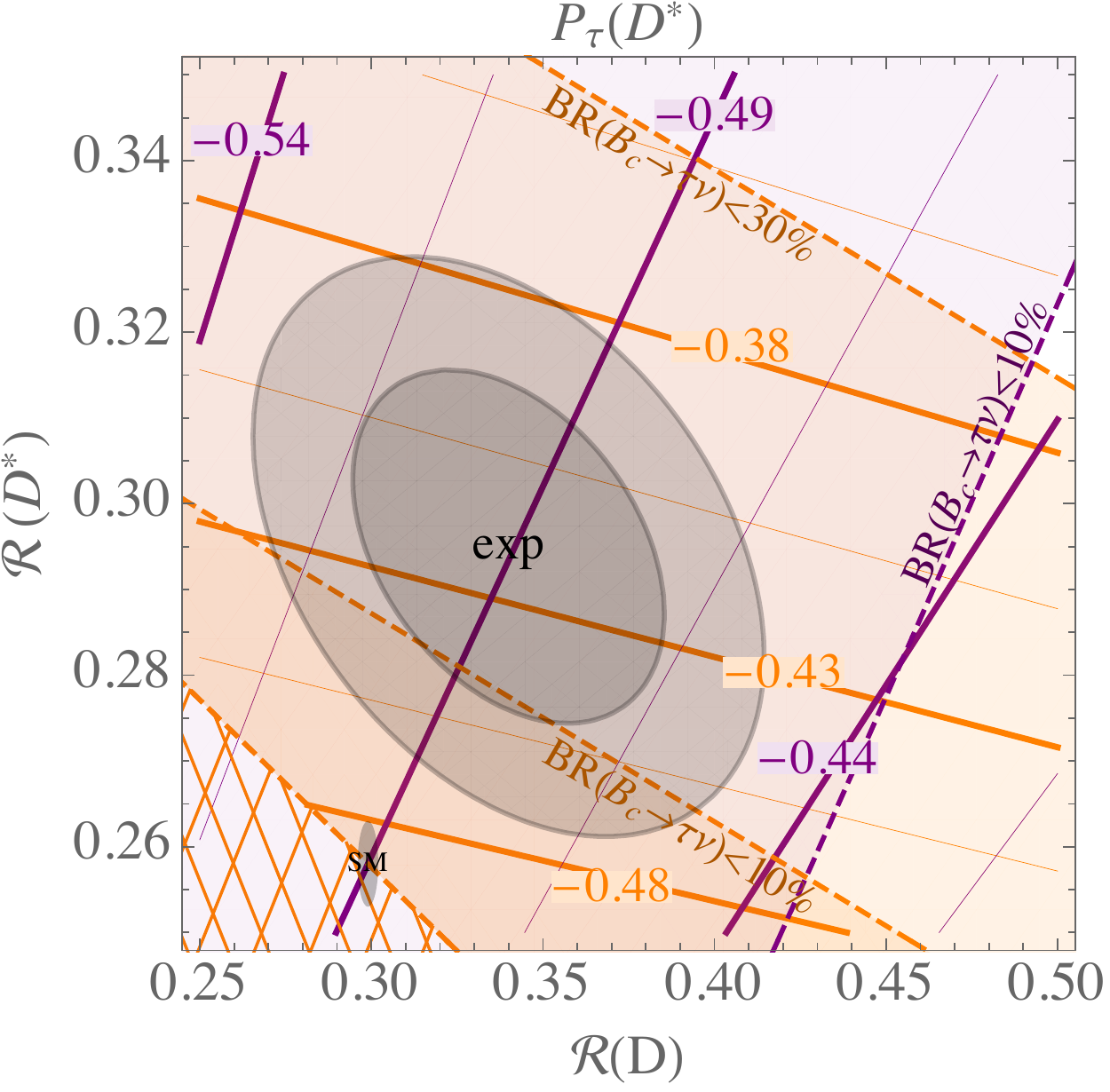}
		}\vspace{-5mm}\\
	\subfigure{	
		\includegraphics[width=0.45\textwidth, bb= 0 0 360 360]{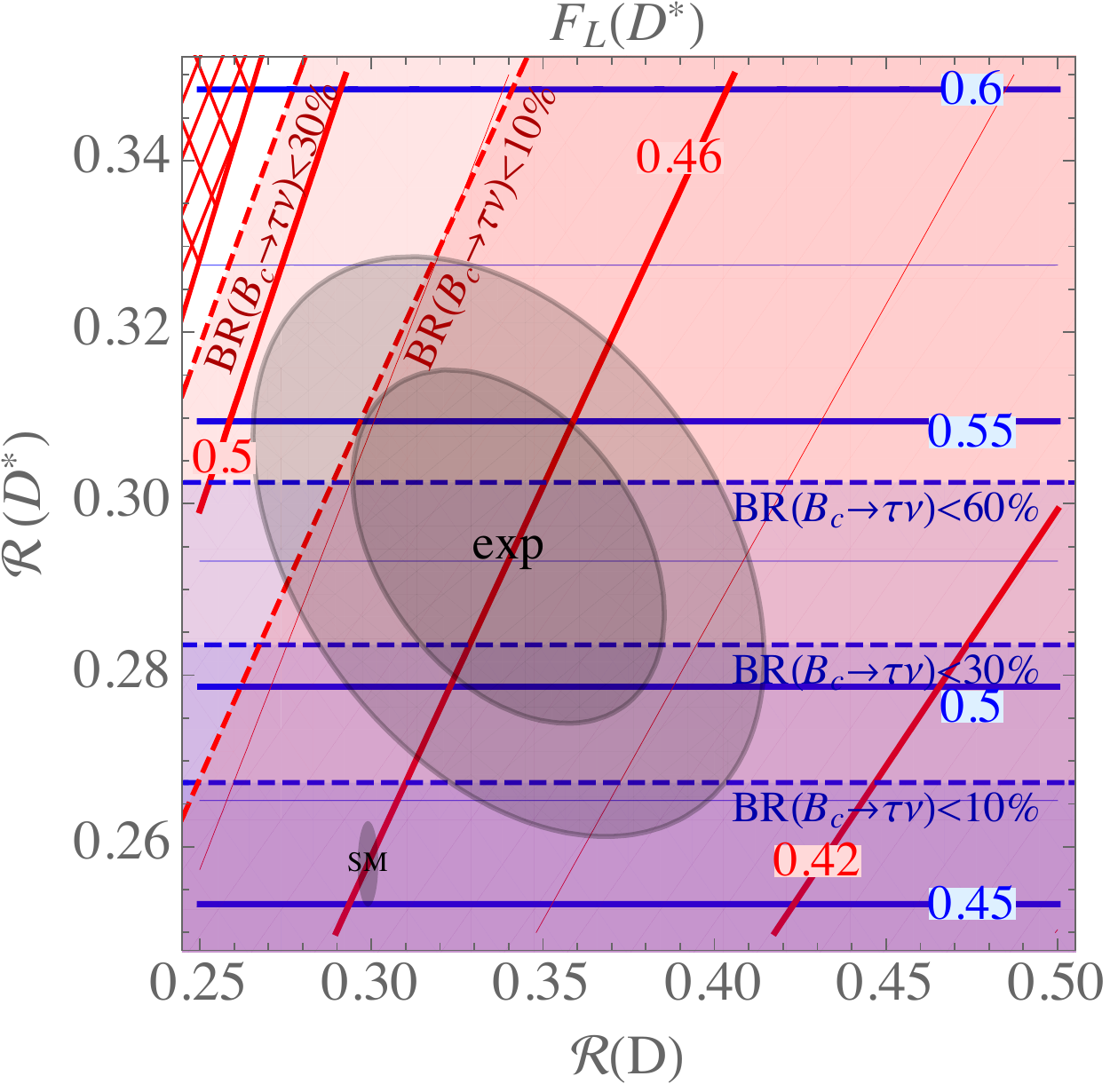}
		}\qquad
	\subfigure{
		\includegraphics[width=0.45\textwidth, bb= 0 0 360 360]{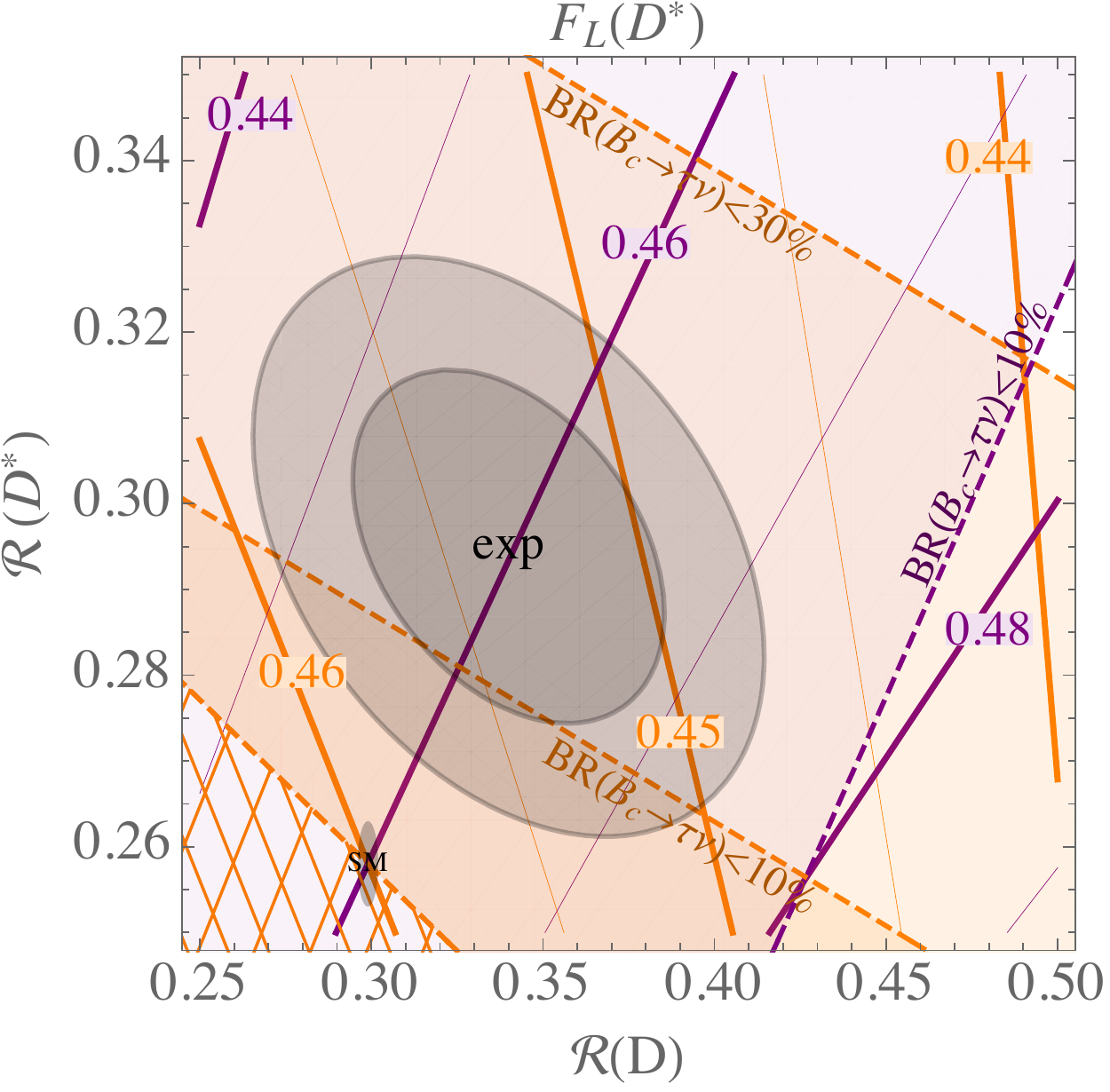}
		}\\	\vspace{1cm}
	\hspace{-5cm}	\includegraphics[width=0.2\textwidth, bb= -55 0 180 38]{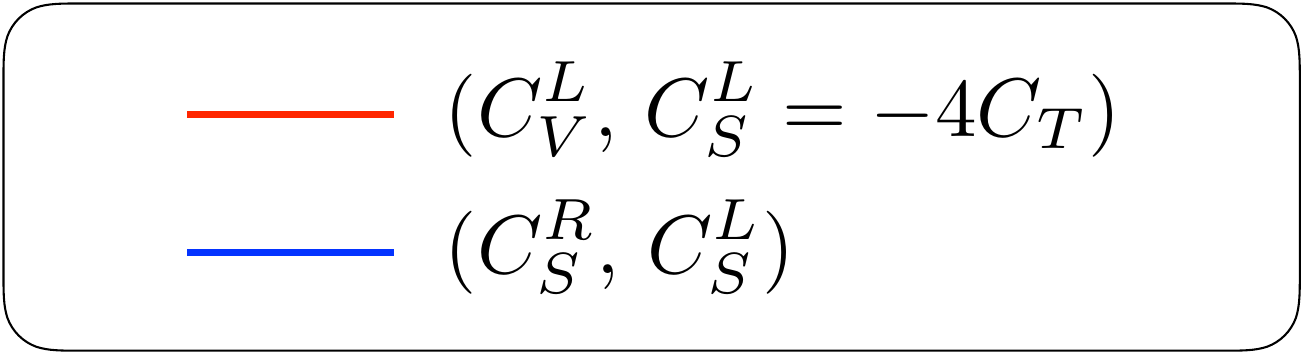}\qquad\qquad\qquad\qquad\qquad\qquad\qquad\qquad
\includegraphics[width=0.1\textwidth,bb = -25 0 95 38]{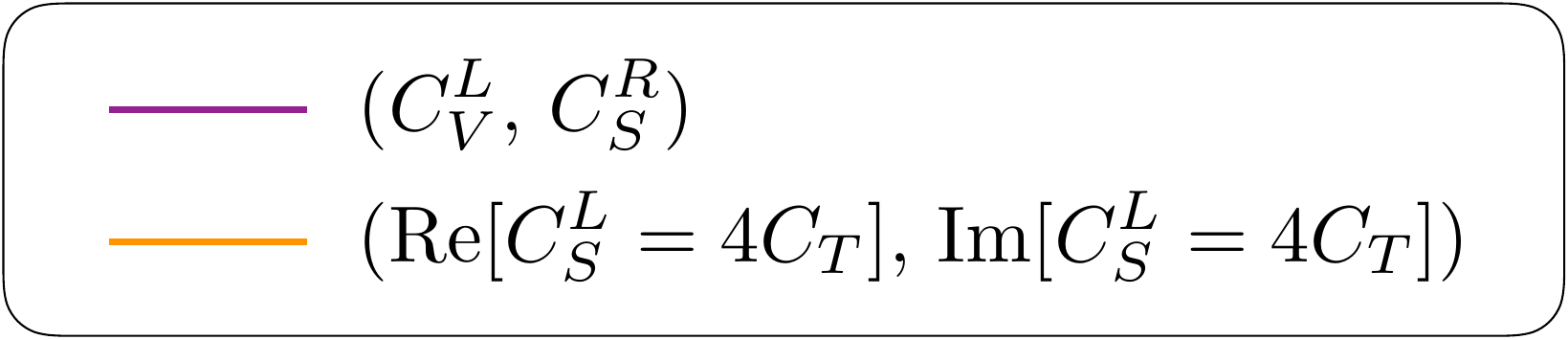}
\end{center}
\vspace{-0.5cm}
\caption{Contour lines of the $\tau$ polarization and the longitudinal
  $D^*$ polarization for the two-dimensional scenarios in the
  ${\cal R}(D)$--${\cal R}(D^*)$ plane, updating Fig.~5 of Ref.~\cite{Blanke:2018yud}.}
\label{RDRDstar1}
\end{figure*}

\begin{figure*}[tp]
\begin{center}
	\subfigure{
		\includegraphics[width=0.45\textwidth, bb=0 0 360 360]{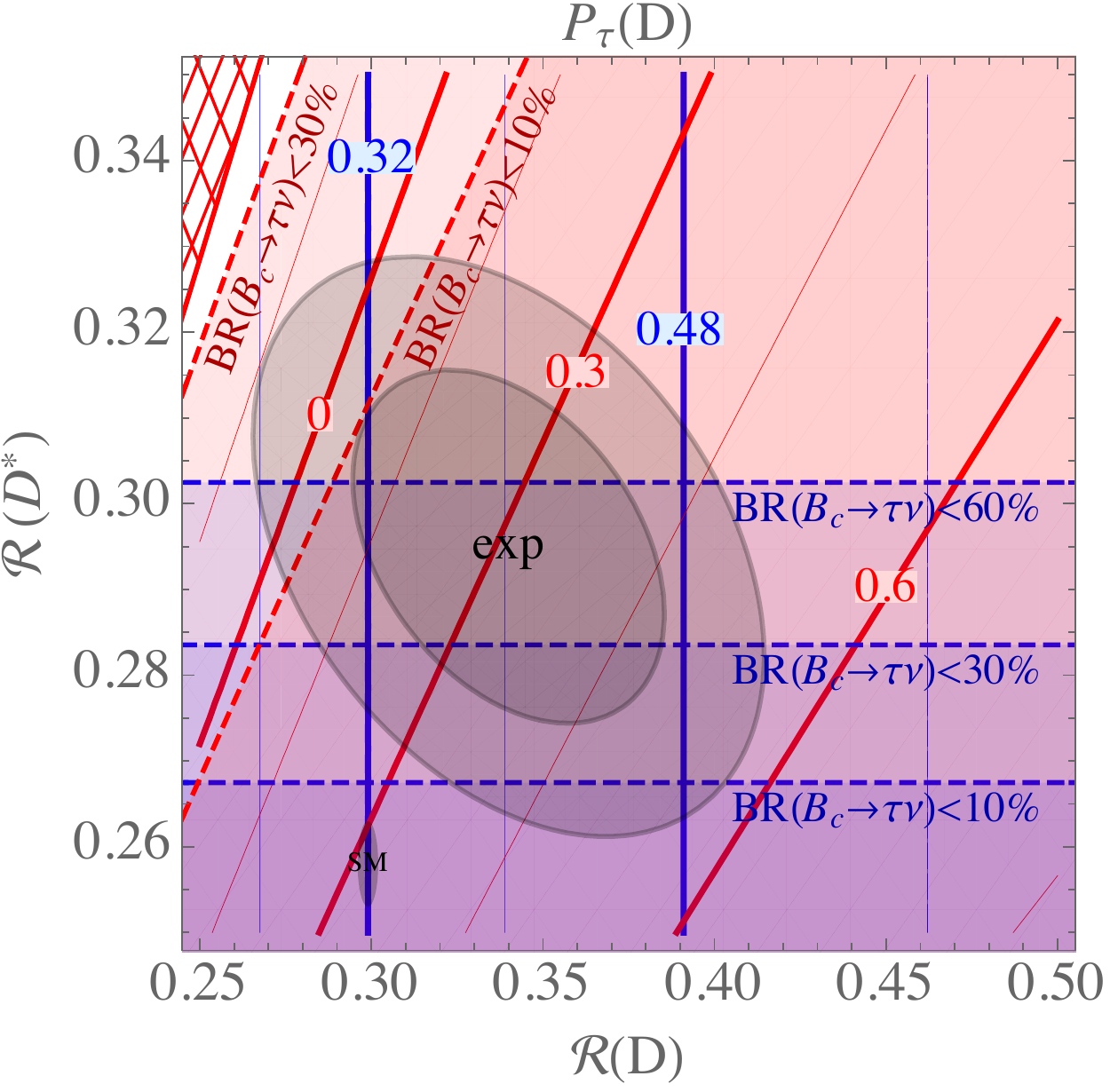}			}\qquad
	\subfigure{		
		\includegraphics[width=0.45\textwidth, bb =0 0  360 360]{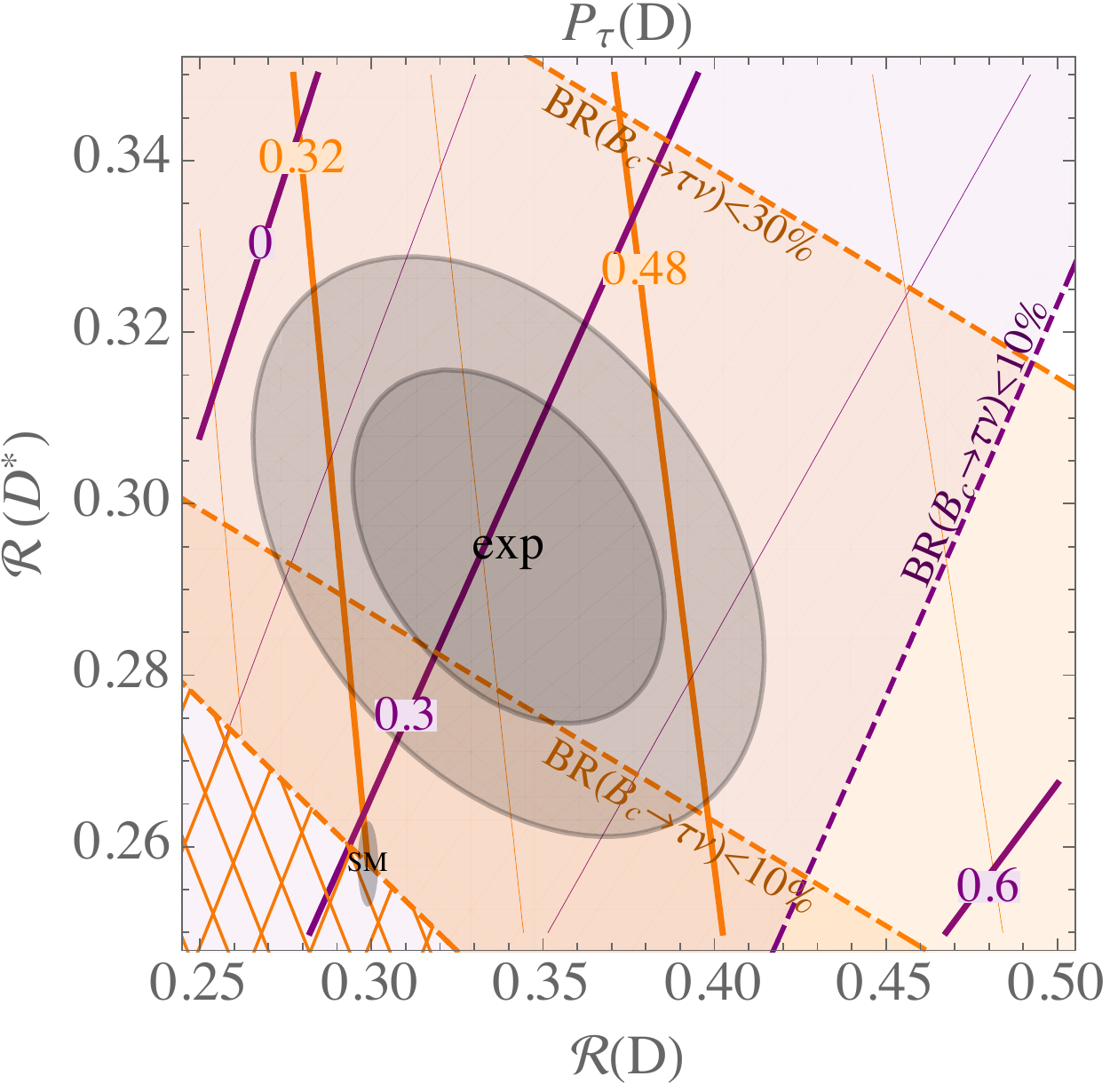}
		}\vspace{-5mm}\\
	\subfigure{	
		\includegraphics[width=0.45\textwidth, bb= 0 0 360 360]{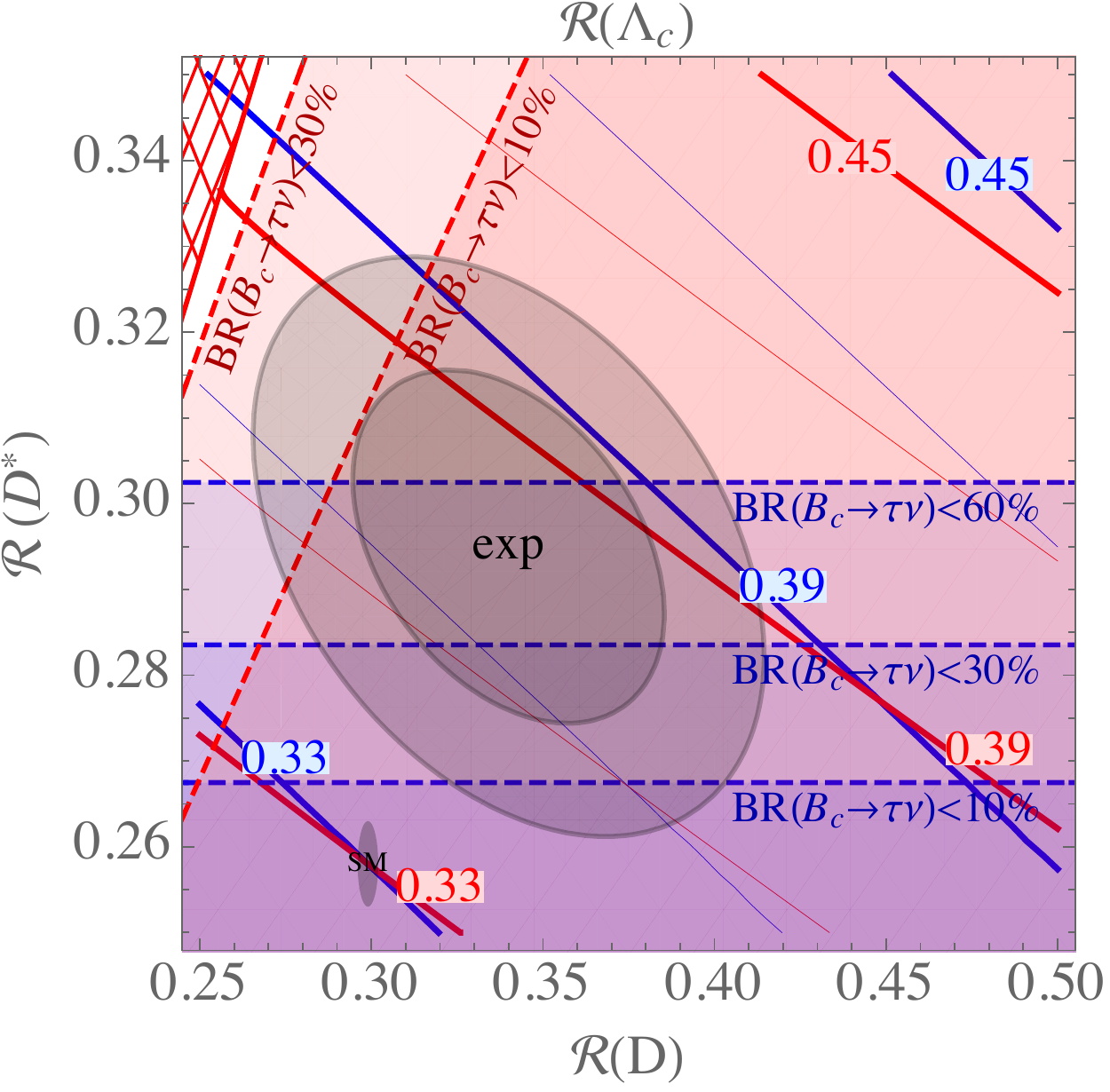}
		}\qquad
	\subfigure{
		\includegraphics[width=0.45\textwidth, bb= 0 0 360 360]{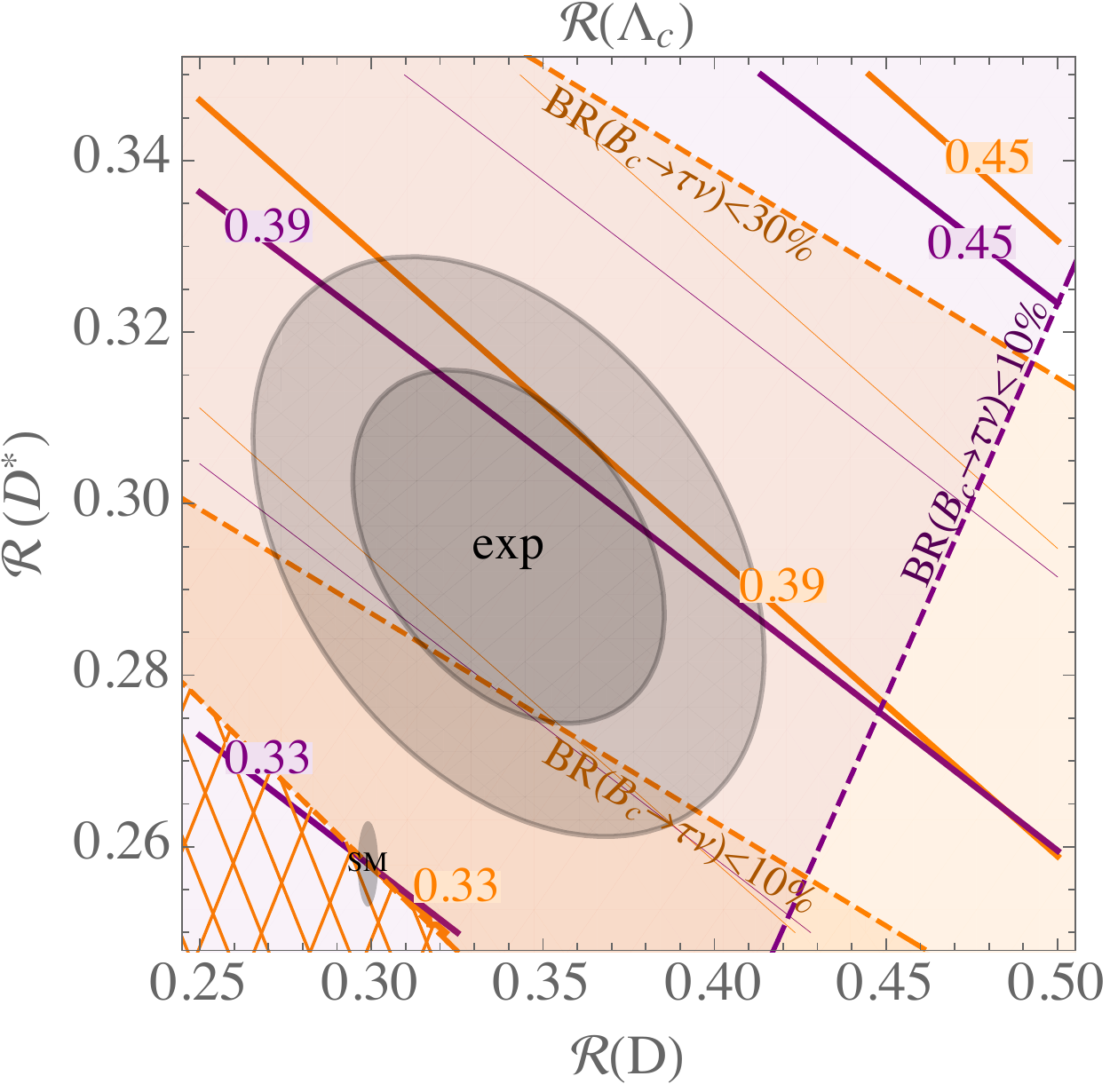}
		}\\	\vspace{1cm}
	\hspace{-5cm}	\includegraphics[width=0.2\textwidth, bb= -55 0 180 38]{legend1.pdf}\qquad\qquad\qquad\qquad\qquad\qquad\qquad\qquad	\includegraphics[width=0.1\textwidth,bb = -25 0 95 38]{legend2.pdf}
\end{center}
\vspace{-0.5cm}
\caption{Contour lines of $P_\tau(D)$ and ${\cal R}(\Lambda_c)$ for the two-dimensional scenarios in the ${\cal R}(D)$--${\cal R}(D^*)$ plane, updating Fig.~6 of Ref.~\cite{Blanke:2018yud}.}
\label{RDRDstar2}
\end{figure*}

In conclusion, 
we {have updated} our fit results for the $b \to c \tau \nu$ anomaly
to include the recent data by the Belle Collaboration~\cite{Abdesselam:2019dgh}.
{The predictions for polarization observables from the fit significantly depend on the Wilson coefficient scenario. Therefore, by accurately probing their correlations at the ongoing Belle II experiment \cite{Kou:2018nap}, one can in principle distinguish between different new physics  models. To exploit their full discriminatory power, however, more precise predictions of the relevant form factors are also necessary.
Furthermore we revisited the constraint on $\bbc$ from LEP data at the $Z$ peak, focusing on the theoretical predictions for the fragmentation of a $b$ quark into a $B_c$ meson, and concluded that our most conservative scenario $\bbc< 60\%$ is not excluded at present.
Moreover, reevaluating our sum rule connecting ${\cal R}(\Lambda_c)$ with ${\cal R}(D^{(*)})$,  we predicted an enhancement of ${\cal R}(\Lambda_c)$ of $(15\pm 4)\%$ with respect to its SM value model-independently, 
which serves as a good experimental cross-check of the $b \to c \tau \nu$ anomaly.}

\medskip

\vspace{2mm} {\it Acknowledgements.}--- {\small 
The work of A.C. is
supported by a Professorship Grant (PP00P2\_176884) of the Swiss National Science Foundation.
The work of 
I.N., and U.N. is supported by the Bundesministerium f\"ur Bildung und Forschung (BMBF, German Federal Ministry of Education and Research) under grant no. 05H18VKKB1. 
The research of M.B. and U.N. is supported by the Deutsche Forschungsgemeinschaft (DFG, German Research Foundation) under grant  396021762 -- TRR 257.
M.M. acknowledges the support of the
DFG-funded Doctoral School ``Karlsruhe School of Elementary and
Astroparticle Physics: Science and Technology''. 
The work of T.K. is supported by Japan Society for the Promotion of Science (JSPS) KAKENHI Grant Number 19K14706.
{We would like to thank 
 Andrew Akeroyd, Debjyoti Bardhan, Giacomo Caria, Shigeki Hirose, {Zoltan Ligeti, Thomas Mannel, and Jure Zupan} for valuable discussions.}


\bibliography{BIB}

\end{document}